\documentclass[final,5p,times,twocolumn]{elsarticle}

\usepackage{graphicx}
\usepackage{subcaption}
\usepackage{xcolor}
\usepackage{amsmath}
\usepackage{amssymb}
\usepackage{bbm}
\usepackage[thinc]{esdiff}
\usepackage{placeins}

\usepackage{array} 
\usepackage{supertabular}
\usepackage{booktabs} 

\newcolumntype{$}{>{\global\let\currentrowstyle\relax}}
\newcolumntype{^}{>{\currentrowstyle}}
\newcommand{\rowstyle}[1]{\gdef\currentrowstyle{#1}%
  #1\ignorespaces
}
\newcolumntype{L}[1]{>{\raggedright\arraybackslash}p{#1}} 
\newcolumntype{C}[1]{>{\centering\arraybackslash}m{#1}} 
\newcolumntype{R}[1]{>{\raggedleft\arraybackslash}p{#1}} 

\DeclareGraphicsExtensions{.pdf,.tikz,.eps,.png}

\definecolor{blueThesis}{rgb}{0,.4,.8} 
\definecolor{darkblueThesis}{rgb}{0.0784314, 0.329412, 0.670588} 
\definecolor{lightblueThesis}{rgb}{0.623529, 0.807843, 1.} 
\definecolor{orangeThesis}{rgb}{.8,.4,.2} 
\definecolor{darkorangeThesis}{rgb}{0.6, 0.3, 0.15} 
\definecolor{lightorangeThesis}{rgb}{1., 0.807843, 0.623529} 
\definecolor{semilightorangeThesis}{rgb}{0.917647, 0.6, 0.462745} 

\renewcommand{\eqref}{\text{Eq. }\ref}
\newcommand{\fig}[1]{Fig. \ref{#1}}
\newcommand{\tab}[1]{Tab. \ref{#1}}

\newcommand\gas{\text{g}}
\newcommand\fm{\text{fm}}
\renewcommand\sf{\text{sf}}
\newcommand\co{\text{co}}
\newcommand\bb{\text{bb}}
\newcommand\+{\oplus}
\renewcommand\-{\ominus}
\newcommand\rad{\text{rad}}
\newcommand\dd{\,\mathrm{d}}
\def\u#1{\,\mathrm{#1}} 
\newcommand{\Kn}{K\!n} 
\newcommand{\Rey}{R\!e} 
\newcommand{\Pe}{P\!e} 
\newcommand{\Pra}{P\!r} 

\def\spaltenvektor#1#2{\left(\begin{array}{#1}#2\end{array}\right)}

\usepackage{natbib}
\bibpunct{(}{)}{;}{a}{}{,}

\journal{Journal of Aerosol Science xxx (2016) xx--–xx}

\begin{document}

\begin{frontmatter}
	
	
	
	\title{Photophoresis on particles hotter/colder than the ambient gas for the entire range of pressures}
	
	
	\author{C. Loesche, T. Husmann}

	\address{Fakult{\"a}t f{\"u}r Physik, Universit{\"a}t Duisburg-Essen, Lotharstr. 1, 47048 Duisburg, Germany}
	
	
	\begin{abstract}
		Small, illuminated aerosol particles embedded in a gas experience a photophoretic force.
		Most approximations assume the mean particle surface temperature to be effectively the gas temperature.
		This might not always be the case.
		If the particle temperature or the thermal radiation field strongly differs from the gas temperature (optically thin gases), given approximations for the free molecule regime overestimate the photophoretic force by an order of magnitude on average and for individual configurations up to three magnitudes.
		We apply the radiative equilibrium condition from the previous paper (Paper 1) --- where photophoresis in the free molecular flow regime was treated --- to the slip flow regime. The slip-flow model accounts for thermal creep, frictional and thermal stress gas slippage and temperature jump at the gas-particle interface. In the limiting case for vanishing Knudsen numbers --- the continuum limit --- our derived formula has a mean error of only 4 \% compared to numerical values.
		Eventually, we propose an equation for photophoretic forces for all Knudsen numbers following the basic idea from Rohatschek by interpolating between the free molecular flow and the continuum limit.		
	\end{abstract}


\end{frontmatter}



\section{INTRODUCTION}
Illuminated particles suspended in a gas experience photophoretic forces \cite{Yalamov1976_photohoresis_fm,Yalamov1976_photohoresis_co, Rohatschek1995, Loesche2013}.
For directed illumination like in \fig{fig:sphere_with_Trot}, a simple description for high Knudsen numbers is based on a kinetic description of the momentum transfer between impinging gas molecules and the particles, which is stronger on one particular side of the particles.
Often this is related to a temperature gradient across the particles' surface which leads to a motion away from the radiation source.

Several experiments show photophoresis \citep{Wurm2008experiments,Loesche2014,vanEymeren2012} and the theoretical treatment of photophoretic forces in different pressure regimes has also progressed \citep{Malai2012AOcOp,Beresnev1993, Yalamov1976_photohoresis_fm, Yalamov1976_photohoresis_co, Reed1977}.

The findings in the first paper \citep{paper1}, (hereafter referred to as Paper 1) are based on work by \cite{Hidy1967,Tong1973,Yalamov1976_photohoresis_fm}, which allow only low radiative fluxes $I$ and small gas-particle temperature differences.
It presented a new free molecular flow (\textit{fm}) approximation, that now also supports the case of considerably higher radiative fluxes ($I$) and hotter/lower surface temperatures with respect to the surrounding gas ($T_\infty$), while assuming the particle to be in equilibrium with an external radiation field at $T_\rad$.
It also performs very well for particles of low thermal conductivity $k$ which so far only \cite{Yalamov1976_photohoresis_fm} does, too.
Paper 1 showed, that the optimized linearizations used have an excellent effect on the results, reducing the minimum and maximum relative error of the analytical equation (within the model) to $\approx\!-50\%$ and 7\%, respectively.

\cite{Beresnev1993,Chernyak1993} proposed an advanced kinetic model for high Knudsen numbers, where also thermal radiation was considered. The external radiation field was at the temperature of the gas. For the \textit{fm} limit they also provide a handy equation, that is similar to the one in Paper 1. However, the model only allowed small radiative fluxes $I$ and therefore only small temperature difference between gas and particle.

Conversely, for low Knudsen numbers, especially in the slip-flow (\textit{sf}) regime, there are hydrodynamic models proposed by \cite{Yalamov1976_photohoresis_co,Reed1977,Mackowski1989aerosol}, where the first work also treats evaporation. None of these models allow high intensities $I$ and also do not account for thermal radiation.
For high intensities and temperature deviance of gas and particle \cite{Malai2012JTePh,Malai2012AOcOp} already proposed a model, incorporating thermal radiation and temperature dependent heat conductivities of gas $k_\gas(T)$ and particle $k(T)$ as well as gas viscosity $\eta(T)$.
Like in \cite{Beresnev1993}, the radiation field is at the temperature of the gas.

In this paper, we apply the findings from Paper 1 on other Knudsen regimes with the aim to find an accurate but handy interpolation function for the entire range of pressures.
This interpolation also supports higher intensities, and therefore the mean particle surface temperature to differ from the gas temperature $T_\infty$.
Furthermore, it also includes the temperature of the radiation field $T_\rad$, which is not necessarily the gas temperature, depending on the setting.
The interpolation is based on approximations for the free molecular \textit{fm} and continuum (\textit{co}) limits following the findings of \cite{Hettner1928,Rohatschek1995}.
However, as we have two temperatures, i.e. $T_\infty$ and $T_\rad$, which do not necessarily have to be the same, we propose another \textit{sf} model in Section \ref{sec:sf_phothophoresis}.
From the equation for the \textit{sf} regime we obtain the limiting case for vanishing Knudsen numbers (\textit{co}).
In the \textit{sf} regime we account for thermal creep, frictional and thermal stress gas slippage and temperature jump at the gas-particle interface.
We will not include temperature dependent $k$, $k_\gas$ and $\eta$ but show how to account for that in Section \ref{sec:discussion}.
For smaller particles the boundary conditions in the \textit{sf} regime can also be extended by some additional addends which are linear in the Knudsen number \citep{Malai2012AOcOp}, introducing several more parameters.
However, as mentioned before, we interpolate between the \textit{co} and \textit{fm} approximations.
Therefore we do not incorporate too many Knudsen-number dependent boundary conditions into this model which vanish in the limiting case $\Kn\to 0$  (\textit{co}).
A discussion of the results and a comparison to other models is done in Section \ref{sec:discussion}.

All variables in this paper are also listed in \tab{tab:notation}, including some basic relations.
Section \ref{sec:supp} provides some additional information for the interested reader in the supplementaries.

\section{CLARIFICATION/KNUDSEN REGIMES}
The Knudsen number $\Kn$ is defined as the ratio of the mean free path of the gas molecules/atoms $\lambda$ and the characteristic length of the problem $r_0$ (here this is the particle radius)
\begin{equation}
	\Kn = \frac{\lambda}{r_0} \; . \label{eq:Knudsenzahl}
\end{equation}
The \textit{fm} and \textit{co} regimes are the limits $\Kn\to\infty$ and $\Kn\to 0$, respectively.
For fixed characteristic particle sizes $r_0$, both limits basically infer $p\to0$ and $p\to\infty$, respectively.
For high Knudsen numbers, the photophoretic force is linear in $p$ (Paper 1).
Conversely, for low Knudsen numbers, the force goes with $p^{-1}$ (this paper).
That means, for both limits it is $\lim\limits_{\Kn\to \infty}F_\text{phot}(p)=\lim\limits_{\Kn\to 0}F_\text{phot}(p)=0$.
This is obviously not useful.
Our considerations made in the \textit{fm} and \textit{co} regimes are hence for large and small enough Knudsen numbers, respectively.

Technically, $\Kn\ge 10$ is associated with the \textit{fm} regime, the transition regime is assumed for a Knudsen number range between $0.25\lesssim\Kn\lesssim 10$, but the lower bound varies with different transfer processes on particles \citep{Hidy1970}.
For low Knudsen numbers $\Kn\ll1$, the \textit{co} regime is extended with a slip-flow boundary condition.
This sub-regime is called the slip-flow regime.
Here, no general bounds can be provided \citep{Hidy1970}.
A sketch of the different regimes is appended in \fig{fig:regimes}.

Therefore it is more exact to say, the considerations made in the \textit{co} regime are actually made in the \textit{sf} regime and only the limiting case for vanishing Knudsen numbers is associated with the \textit{co} regime.
On the other hand, as \textit{fm} photophoresis is not meant for zero pressure  ($\Kn \neq 0$), one can also talk about \textit{co} photophoresis ($\Kn \neq \infty$).


\section{PHOTOPHORESIS AT LOW KNUDSEN NUMBERS}\label{sec:sf_phothophoresis}
\begin{figure}[ht!]
	\centering
	\includegraphics{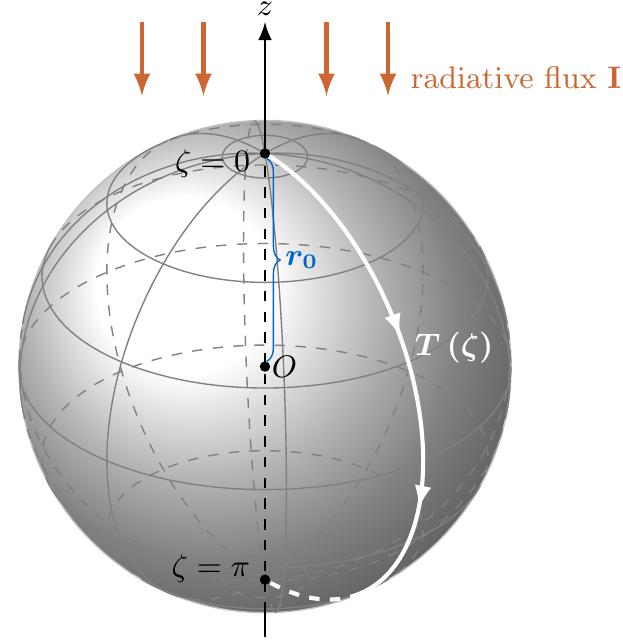}
	\caption{\label{fig:sphere_with_Trot}
		Visualization of the situation considered.
		Illumination is directed along $z$-axis, thus for a homogeneous particle the surface temperature only depends on $\zeta$ (spherical coordinate system ($r,\zeta,\xi$)).
		The sphere's radius is $r_0$.
		The temperature of the gas is $T_\infty$ ($r\to\infty$), the temperature of the radiation field is $T_\rad$.
	}	
\end{figure}
For solid particles at low Knudsen numbers (e.g. large aerosols) the photophoretic force is a direct result of thermal creep along a surface $\partial V$ of the suspended particle \citep{Reed1977,Bakanov2004}, which occurs in case of a temperature gradient in the gas, which is tangential to $\partial V$. 

For directed illumination of a homogeneous, spherical particle embedded in an effectively infinite gas as shown in \fig{fig:sphere_with_Trot}, an equation for the ensuing \textit{longitudinal} photophoretic force at low Knudsen numbers is proposed.
The particle is supposed to be in a radiative equilibrium with an external radiation field at temperature $T_\rad$.
This radiation field can also be emitted by the gas itself, which has the temperature $T_\infty$ far away from the suspended particle.
We present two means to describe photophoresis for directed illumination at a radiative flux of $I$.
One is solely for the slip-flow regime with the limiting case of $\Kn \rightarrow \infty$ (\textit{co}) and the second one interpolates between all regimes, using the \textit{co} limit and the \textit{fm} limit from Paper 1.

The model consists of a hydrodynamic part and a heat transfer part.
In this setting (\fig{fig:sphere_with_Trot}), both problems are axisymmetric in $\xi$, i.e. they only depend on the coordinates $r$ and $\zeta$.
The $z$-axis is therefore set parallel to the direction of illumination and motion at speed $u$, and especially: $\mathbf{e}_z = -\mathbf{e}_I$.
Gases and fluids with a small dynamic viscosity can be treated as ideal fluids.
Additionally, if the fluid is incompressible and the flow is free of vortices, the flow can be treated like a potential flow.
However, this statement is right for almost every point in the fluid except at the particle-fluid interface.
Friction will definitely contribute here, large flow speed gradients occur, and friction forces will be comparable to inertial forces.
Therefore, the boundary conditions in this model account for thermal creep as well as frictional and thermal stress gas slippage at the gas-particle interface.

Before setting up the hydrodynamic model, we give a short insight into thermal creep.

\subsection{Thermal creep}
Thermal creep causes a gas flow tangential to a surface (tangent $\mathbf{t}$, normal $\mathbf{n}$) at a mass speed $v$ which obeys the equation \citep{Brenner2009}
\begin{equation}
	\left(\mathbbm{1}-\mathbf{n}\otimes\mathbf{n}\right)\cdot\left(\mathbf{v}-\mathbf{u}\right) = \kappa_\text{s} \, \eta_\text{kin}\left(\mathbbm{1}-\mathbf{n}\otimes\mathbf{n}\right)\cdot\boldsymbol{\nabla} \log T_\gas \qquad\text{on }\partial V\; . \label{eq:thermal_creep_velocity}
\end{equation}
The mass velocity $\mathbf{v}$ obeys the continuity equation
\begin{equation}
	\frac{\partial \rho}{\partial t}+\boldsymbol{\nabla}\cdot\rho\, \mathbf{v} = 0 \qquad\text{on }\partial V \; ,
\end{equation}
$\mathbf{u}$ is the velocity of the surface $\partial V$ relative to the gas, $\eta_\text{kin}$ denotes the kinematic viscosity of the gas, and $\rho$ and $T_\gas$ the gas mass density and 
gas temperature, respectively \citep{Brenner2005}.
$\kappa_\text{s}$ is the thermal creep coefficient (also thermal slip coefficient) \footnote{\cite{Brenner2006,Brenner2009} also proposed a nonmolecular thermodynamic theory of thermal creep, based on a hydrodynamic theory, that is valid for physiochemically and thermally inert solids suspended in not only gases but also fluids as
	\begin{equation*}
	\left(\mathbbm{1}-\mathbf{n}\otimes\mathbf{n}\right)\cdot\left(\mathbf{v}_\mathrm{m}-\mathbf{u}\right) = D \, \gamma_\text{exp}\left(\mathbbm{1}-\mathbf{n}\otimes\mathbf{n}\right)\cdot\boldsymbol{\nabla} T_\gas \qquad\text{on }\partial V \; , 
	\end{equation*}
	introducing the fluid's self-diffusion coefficient $D$ and the fluid's thermal expansion coefficient (at constant pressure) $\gamma_\text{exp} = -\frac1{\rho}\left(\frac{\partial\rho}{\partial T_\gas}\right)_p$ \; . \label{eq:thermal_creep}
	In contrast to this equation, \eqref{eq:thermal_creep_velocity} is only valid for gases and no restrictions on the solids are imposed.}.
\cite{Brenner2009} points out, that various experts on molecular dynamics agree on the correctness of this equation for gases, even though the underlying gas-kinetic molecular theory is not rigorous but only semi-quantitative.

The original value of the thermal creep coefficient $\kappa_\text{s}=\frac34$ goes back to \cite{Maxwell1879}.
\cite{Bakanov1992} lists a couple of parameters $a_{\kappa_{\text{s}}}$ and $b_{\kappa_{\text{s}}}$ for different models which relate $\kappa_\text{s}$ and the momentum accommodation coefficient $\alpha_\text{m}$ by the equation
\begin{equation}
	\kappa_\text{s}(\alpha_\text{m}) \simeq \frac34\left(a_{\kappa_{\text{s}}}+b_{\kappa_{\text{s}}}\,\alpha_\text{m}\right) \; , \label{eq:thermal_slip_coeff}
\end{equation}
where $a_{\kappa_{\text{s}}}$ is close to 1 and $b_{\kappa_{\text{s}}}$ around 0.5, thus the thermal creep coefficient can be expected to obtain values between $0.75\le\kappa_\text{s}\le 1.24$.
\cite{Rohatschek1995} assumes a value of $\kappa_\text{s}=1.14$ for $\alpha_\text{m}=0.9$ and this value is also used by \cite{Loesche2014,Hesse2011Diplom}.
\cite{Ivchenko1993} also suggested a model with more accurate values for $\kappa_\text{s}$. One of the latest works is \cite{Ivchenko2007}.

\subsection{Hydrodynamic model}
The momentum balance in the fluid is given by
\begin{equation}
	\rho\frac{\mathrm{d}\mathbf{v}}{\mathrm{d}t} \equiv \rho\left(\frac{\partial\mathbf{v}}{\partial t} + \left(\mathbf{v}\cdot\boldsymbol{\nabla}\right)\mathbf{v}\right) = -\boldsymbol{\nabla}p + \boldsymbol{\nabla}\cdot\underline{\boldsymbol{\sigma}} + \rho\,\mathbf{F}_\text{ext} \; , \label{eq:momentum_balance}
\end{equation}
where $\underline{\boldsymbol{\sigma}}$ denotes the stress tensor, that is related to the friction tensor $\underline{\boldsymbol{R}}$ \footnote{Inserting \eqref{eq:sigma} into \eqref{eq:momentum_balance} yields the Navier-Stokes equation \begin{equation*}
		\rho\left(\frac{\partial\mathbf{v}}{\partial t}+\left(\mathbf{v}\cdot\boldsymbol{\nabla}\right)\mathbf{v}\right) = -\boldsymbol{\nabla}p + \eta_\text{dyn}\Delta\mathbf{v} + \frac13\eta_\text{dyn}\boldsymbol{\nabla}\left(\boldsymbol{\nabla}\cdot\mathbf{v}\right) + \rho\,\mathbf{F}_\text{ext} \; .
	\end{equation*}}
\begin{subequations}
	\begin{align}
		\sigma_{ik} &= -p\,\delta_{ik} + R_{ik} \\
		R_{ik} &= \eta_\text{dyn}\left(\frac{\partial v_i}{\partial x_k}+\frac{\partial v_k}{\partial x_i}\right) - \delta_{ik} \frac23 \eta_\text{dyn} \boldsymbol{\nabla}\cdot\mathbf{v} \; .
	\end{align}
	\label{eq:sigma}
\end{subequations}
The photophoretic motion of the particle causes the gas to move at small Reynolds numbers \Rey, hence the convective acceleration $\left(\mathbf{v}\cdot\boldsymbol{\nabla}\right)\mathbf{v}$ can be omitted (vortex-free fluid: $\boldsymbol{\nabla}\times\mathbf{v} = 0$).
Furthermore, we want to get the quasi-stationary solution ($\partial_t\mathbf{v}=0$) for the incompressible fluid (source-free velocity field $\boldsymbol{\nabla}\cdot\mathbf{v} = 0\Rightarrow\partial_t\rho=0$).
Eventually, we have no body force ($\mathbf{F}_\text{ext} = 0$).
Therefore \eqref{eq:momentum_balance} simplifies to
\begin{equation}
	\boldsymbol{\nabla}p = \eta_\text{dyn}\,\Delta\mathbf{v} \; . \label{eq:momentum_balance2}
\end{equation}
Because of $\boldsymbol{\nabla}\times\mathbf{v}=0$, $\mathbf{v}$ has a scalar potential ($\boldsymbol{\nabla}\times\boldsymbol{\nabla} f = 0$ for a scalar function $f$).
Also, because of $\boldsymbol{\nabla}\cdot\mathbf{v}=0$, $\mathbf{v}$ has a vector potential, generally written as $\boldsymbol{\Psi}$.

Considering the symmetry of the three-dimensional problem, the fluid/gas velocity is
\begin{equation}
\mathbf{v} = v_r \, \mathbf{e}_r + v_\zeta \, \mathbf{e}_\zeta \; , \label{eq:fluid_velocity}
\end{equation}
and therefore quasi-two-dimensional.

\subsubsection{Ansatz}
In orthogonal coordinates $q_1,q_2,q_3$ (with the accompanying scaling factors $h_1,h_2,h_3$) a three-dimensional, stationary flow of an incompressible Newtonian fluid with symmetry in $q_3$ has a vector potential that only depends on two variables $\boldsymbol{\Psi}=\boldsymbol{\Psi}(q_1,q_2)$.
Therefore it can be set $\boldsymbol{\Psi}\sim\psi\mathbf{e}_3$, and the
velocity can subsequently be written as
\begin{subequations}
\begin{align}
   \mathbf{v} &= \boldsymbol{\nabla}\times\boldsymbol{\Psi} \\ &=-\boldsymbol{\nabla}\times\left(\psi\frac{\mathbf{e}_3}{h_3}\right) \\
   &= \frac{\mathbf{e}_3}{h_3}\times\mathbf{\nabla}\psi(q_1,q_2) -\psi(q_1,q_2)\mathbf{\nabla}\times\frac{\mathbf{e}_3}{h_3} \\
   &= \frac{\mathbf{e}_3}{h_3}\times\mathbf{\nabla}\psi(q_1,q_2) \; . \label{eq:v}
\end{align}
\end{subequations}
$\psi$ is called the Stokes stream function.
Applying $\mathbf{e}_3\times\mathbf{\nabla}$ on \eqref{eq:momentum_balance2} (and using \eqref{eq:v}
) yields the equation that $\psi$ satisfies, that is also the governing equation for the flow \citep{Schubert2015Geophysics}
\begin{subequations}
	\begin{align}
		E^4\,\psi &= 0 \\
		E^2 &= \frac{h_3}{h_1 h_2}\left[\frac{\partial}{\partial q_1}\left(\frac{h_2}{h_1 h_3}\frac{\partial}{\partial q_1}\right) + \frac{\partial}{\partial q_2}\left(\frac{h_1}{h_2 h_3}\frac{\partial}{\partial q_2}\right)\right] \label{eq:stream_function_operator} \; .
	\end{align}
	\label{eq:stream}
\end{subequations}
In spherical coordinates $(r,\zeta,\xi)$ the scaling factors are $(h_1,h_2,h_3)=(1,r,r\,\sin\zeta)$, and hence it is $E^2 = \partial_{r,r} + \frac{\sin\zeta}{r^2} \partial_\zeta\left(\frac1{\sin\zeta}\partial_\zeta\right)$.
The velocity subsequently reads
\begin{equation}
   \mathbf{v} = \spaltenvektor{c}{ v_r \\ v_\zeta \\ v_\xi } = \frac1{r\,\sin\zeta} \spaltenvektor{c}{ -\frac1{r}\,\partial_\zeta\psi \\ \partial_r\psi \\ 0 } \; .
   \label{eq:v_r_and_zeta}
\end{equation}
The ansatz for $\psi$ is \citep{Reed1977}
\begin{subequations}
\begin{align}
   \psi(r,\zeta) &= u\,\psi_R(r)\, \psi_Z(\zeta)\\
   \psi_Z(\zeta) &= \frac12 \sin^2\zeta \; .
\end{align}
\label{eq:stream_function}
\end{subequations}
The radial part $\psi_R(r)$ is determined by the governing equation $E^4\,\psi = 0$, which formulates an ordinary differential equation for $\psi_R(r)$.
Its solution is
\begin{equation}
   \psi_R(r) = \frac{a}{r}+b\,r+c\,r^2+d\,r^4 \; .
\end{equation}
In the following, the gas temperature is expanded into a Legendre series
\begin{align}
   T_\gas(r,\zeta) &= T_\infty + \sum\limits_{\nu=0}^{\infty}C_\nu \left(\frac{r_0}{r}\right)^{\nu+1} P_\nu(\cos\zeta) \label{eq:T_g} \; .
\end{align}

\subsubsection{Boundary conditions}
Like in \cite{Reed1977}, we use an inertial reference frame at rest with the fluid far away from the particle (\eqref{eq:bc3}), where the $z$-axis is parallel to the direction of illumination (due to symmetry in $\xi$, see \fig{fig:sphere_with_Trot}).
The fluid does not penetrate the particles surface, therefore the fluid velocity has no additional normal component than $u\,\cos\zeta$ (\eqref{eq:bc1}).
The thermal creep introduces a (tangential) boundary condition with symmetry in $\xi$, given by \eqref{eq:thermal_creep}.
To be able to use orthogonality relations, \eqref{eq:thermal_creep} is linearized at the mean near-surface temperature of the gas \footnote{$\left(\mathbbm{1}-\mathbf{n}\otimes\mathbf{n}\right)\cdot\mathbf{u} = \spaltenvektor{c}{ 0 \\ u_\zeta \\ u_\xi }$ with $u_\zeta=-u\,\sin\zeta$ and $u_\xi=0$ is separately put in the boundary condition \eqref{eq:bc2} as not only $\mathbf{u}$ but other addends occur, too.} ($\eta_\text{kin}=\eta_\text{dyn}/\rho$)
\begin{equation}
	v_\mathbf{t} \simeq \left.\kappa_\text{s} \, \frac{\eta_\text{dyn}}{\rho\,r_0\,\overline{\left.T_\gas\right|_{\partial V}}}\frac{\partial T_\gas}{\partial_\zeta}\right|_{r=r_0} \; ,
\end{equation}
where $\left.\frac{\partial T_\gas}{\partial \mathbf{t}}\right|_{\partial V} = \left.\frac1{r_0}\frac{\partial T_\gas}{\partial_\zeta}\right|_{r=r_0}$.
As the friction forces are strong at the particle-gas interface, we account for shear stress, in spherical coordinates given as
\begin{equation}
   R_{\zeta r} = \eta_\text{dyn}\left(\frac1{r}\frac{\partial v_r}{\partial\zeta}+r\frac{\partial}{\partial r}\left(\frac{v_\zeta}{r}\right)\right) \; , \label{eq:Rzetar}
\end{equation}
and thermal stress \citep{Chang2012_thermoPhotophoresis}
\begin{equation}
   \sigma_\text{t} = -\frac{\eta_\text{dyn}^2}{\rho\,\overline{\left.T_\gas\right|_{\partial V}}}\left(\frac{1}{r}\frac{\partial^2 T}{\partial r\,\partial\zeta} -\frac1{r^2}\frac{\partial T}{\partial\zeta}\right) \; .
\end{equation}
Summarizing, the boundary conditions are given as \citep{Reed1977,Chang2012_thermoPhotophoresis}
\begin{subequations}
\begin{align}
	v_r &= u\,\cos\zeta \qquad\text{on }\partial V \label{eq:bc1} \\
	v_\zeta &= -u\,\sin\zeta + v_\mathbf{t} + \frac{\kappa_\text{m}\,\Kn\,r_0}{\eta_\text{dyn}}\left(\sigma_{\zeta r}+\kappa_\text{h}\,\sigma_\text{t}\right) \qquad\text{on }\partial V \label{eq:bc2} \\
	v_r &\xrightarrow{r\rightarrow\infty} 0\; ,\qquad v_\zeta\xrightarrow{r\rightarrow\infty} 0  \label{eq:bc3} \; .
\end{align}
\end{subequations}
The values for the thermal stress slip coefficient $\kappa_\text{h}$ vary between 1 and 3 \citep{Chang2012_thermoPhotophoresis}; $\kappa_\text{m}$ is the gas-kinetic frictional slip which is related to the momentum accommodation coefficient $\alpha_\text{m}$, with values around $1.00 \le \kappa_\text{m} \le 1.35$ and typically taking about $1.25$ \citep{Reed1977}.

\subsubsection{Solution}
The velocity $\mathbf{v}=v_r\mathbf{e}_r + v_\zeta\mathbf{e}_\zeta$ is completely given by Eqs. \ref{eq:v_r_and_zeta}--\ref{eq:T_g},
and the unknown parameters $a,b,c$ and $d$ are restricted by the boundary conditions.
From \eqref{eq:bc3}, it can be concluded that
\begin{equation}
   c=d=0 \; .
\end{equation}
Because of \eqref{eq:bc1}, it is 
\begin{equation}
   a+r_0^2 \left(b+r_0\right)=0 \; . \label{eq:bc1_res}
\end{equation}
\eqref{eq:bc2} involves a derivation of the Legendre series of the gas temperature in $\zeta$ (\eqref{eq:T_g}). As it is $\partial_\zeta P_\nu(\cos\zeta) = P_\nu^1(\cos\zeta)$, only the polynomials $P_\nu^1$ occur in \eqref{eq:bc2}.
All terms with $u$ and $a,b$ are linear in $\sin\zeta=P_1^1(\cos\zeta)$.
As pairwise different $P_\nu^1$ are orthogonal to each other (see \eqref{eq:orthogonality} in the appendix for details), a scalar product of this boundary condition with $P_\nu^1$ will isolate the interesting addends containing $a$ and $b$.
Together with \eqref{eq:bc1_res} it is
\begin{subequations}
\begin{align}
	a &= \frac{r_0^3}{1+3 \kappa_\text{m} \Kn}\left(\frac12+\left(\kappa_\text{s}+3 \kappa_\text{h} \kappa_\text{m} \Kn \right)\frac{\eta_\text{dyn} }{\rho\,r_0\,\overline{\left.T_\gas\right|_{\partial V}}\,u}\,C_1\right) \\
	b &= -\frac{r_0}{1+3 \kappa_\text{m} \Kn}\left(\frac12+\left(\kappa_\text{s}+3 \kappa_\text{h} \kappa_\text{m} \Kn \right)\frac{\eta_\text{dyn} }{\rho\,r_0\,\overline{\left.T_\gas\right|_{\partial V}}\,u}\,C_1\right) -r_0 \; ,
\end{align}
\end{subequations}
and subsequently
\begin{subequations}
	\begin{align}
		v_r &= \frac{\cos\zeta\,r_0^2}{2\rho\,\overline{\left.T_\gas\right|_{\partial V}}\,r^3} \frac{2 C_1 \eta_\text{dyn}  \left(1+\frac{r^2}{r_0^2}\right) \left(\kappa_\text{s}+3 \kappa_\text{h} \kappa_\text{m} \Kn\right)}{1 + 3 \kappa_\text{m} \Kn} \cdot \\
		&\quad\cdot \frac{\rho\,r_0\,\overline{\left.T_\gas\right|_{\partial V}}\,u \left(1+3 \frac{r^2}{r_0^2} \left(1+2 \kappa_\text{m} \Kn\right)\right)}{1 + 3 \kappa_\text{m} \Kn} \\
		v_\zeta &= \frac{\sin\zeta\,r_0^2}{4\rho\,\overline{\left.T_\gas\right|_{\partial V}}\,r^3} \frac{2 C_1 \eta_\text{dyn}  \left(1-\frac{r^2}{r_0^2}\right) \left(\kappa_\text{s}+3 \kappa_\text{h} \kappa_\text{m} \Kn\right)}{1 + 3 \kappa_\text{m} \Kn} \cdot \\
		&\quad\cdot \frac{\rho\,r_0\,\overline{\left.T_\gas\right|_{\partial V}}\,u \left(1-3 \frac{r^2}{r_0^2} \left(1+2 \kappa_\text{m} \Kn\right)\right)}{1 + 3 \kappa_\text{m} \Kn} \; .
	\end{align}
\end{subequations}

Here, due to the symmetry of the setting, only $F_z$ is not zero \citep{Happel1983}:
\begin{align}
   F_z &= -8\pi\,\eta_\text{dyn}\lim\limits_{r\to\infty}\frac{r\,\psi(r,\zeta)}{r^2\sin^2\zeta}\\
   &= -4\pi\,\eta_\text{dyn}\,u\,b \; . \label{eq:Fz}
\end{align}
Inserting $v_r$ and $v_\zeta$ into \eqref{eq:Fz} yields the force as
\begin{equation}
   F_z = -4\pi \frac{\eta^2_\text{dyn} }{ \rho\,\overline{\left.T_\gas\right|_{\partial V}} } \frac{\kappa_\text{s}+3 \kappa_\text{h} \kappa_\text{m} \Kn}{1+3 \kappa_\text{m} \Kn}\,C_1 - 6 \pi\,\eta_\text{dyn}\,r_0\,u\,\frac{1+2 \kappa_\text{m} \Kn}{1+3 \kappa_\text{m} \Kn} \; .
\end{equation}
In the steady state, where the particle moves at constant speed $u$, it is $F_z=0$.
That means, two forces are compensate each other, that is the photophoretic force
\begin{equation}
	\mathbf{F}_\text{phot} = -4\pi\frac{\eta^2_\text{dyn} }{ \rho\,\overline{\left.T_\gas\right|_{\partial V}} } \frac{\kappa_\text{s}+3 \kappa_\text{h} \kappa_\text{m} \Kn}{1+3 \kappa_\text{m} \Kn}\,C_1 \, \mathbf{e}_z \label{eq:Fphot}
\end{equation}
and the drag/resistance force
\begin{equation}
\mathbf{F}_\text{drag} = -6\pi\,\eta_\text{dyn}\,r_0\, \frac{1+2\kappa_\text{m} \Kn}{1+3\kappa_\text{m} \Kn} \,\mathbf{u}_\text{phot} \; .
\end{equation}
The ensuing steady state velocity is
\begin{align}
   \mathbf{u}_\text{phot} &= -\frac23\frac{\eta_\text{dyn} }{ \rho\,\overline{\left.T_\gas\right|_{\partial V}}\,r_0 } \frac{\kappa_\text{s}+3 \kappa_\text{h} \kappa_\text{m} \Kn}{1+2 \kappa_\text{m} \Kn}\,C_1 \, \mathbf{e}_z \; . \label{eq:uphot}
\end{align}
Instead of this equation, the Millikan drag equation can be used here, which is more accurate for $\Kn\approx1$ \citep{Mackowski1989aerosol}.

In the following section, the unknown expansion coefficient $C_1$ of the gas temperature is determined by solving a heat transfer problem.

\subsection{Heat transfer model}
We follow the assumptions made in Paper 1, the particle is heated by directed illumination, which is described by the inhomogeneity $I\, q(r,\cos\zeta)$ in the heat transfer equation.
The heat transfer model supports energy exchange with the gas, thermal radiation and a temperature jump at the gas-particle interface.
The gas is supposed to be at temperature $T_\infty$ far away from the suspended particle.
The particle is required to be in radiative equilibrium with an external radiation field at temperature $T_\rad$.
This can also be the gas itself as $T_\rad=T_\infty$.

\subsubsection{Ansatz}
The P{\'e}clet number $\Pe$ (\eqref{eq:Pe}) is required to be small, then thermal diffusive transport predominates advective transport.
Therefore the governing equations are
\begin{subequations}
\begin{align}
	k\,\Delta T &= - I\, q(r,\cos\zeta) \label{eq:heatEQ1} \\
	k_\gas\,\Delta T_\gas &= 0 \; ,\label{eq:heatEQ2}
\end{align}
\label{eq:heatEQ}
\end{subequations}
for the particle and gas, respectively.
$I=\varepsilon\,I_0$ is the absorbed radiative flux, $\varepsilon$ denotes the emissivity \footnote{In standard form $q(r,\zeta)$ is \citep{Yalamov1976_photohoresis_co,Malai2012JTePh} \begin{align*}
	q(r,\zeta) &= 2\chi_1\,\chi_2\, k_0 B(r,\zeta) \\
	B(r,\zeta) &= \frac1{2\pi}\int\limits_{0}^{2\pi}\frac{|E(r,\zeta,\xi)|}{E_0^2}\dd\xi \; ,
	\end{align*} where $\chi=\chi_1+\imath\,\chi_2$ is the complex refractive index and $k_0$ the wave number of an electromagnetic wave at amplitude $E_0$.}.

The ansatz for the particle temperature $T$ is constructed insofar that on the surface it is given by the simple equation
\begin{align}
	T(r_0,\zeta) &= \sum\limits_{\nu=0}^{\infty} A_\nu \, P_\nu(\cos\zeta) \; . \label{eq:newApproximation_Tsurface}
\end{align}
For the general solution $T(r,\zeta)=T_1(r,\zeta)+T_2(r,\zeta)$, the homogeneous and particular ansatz functions are
\begin{subequations}
	\begin{align}
	T_1(r,\zeta) &= \sum\limits_{\nu=0}^{\infty}\left(A_\nu-B_\nu\,J_\nu(r_0)\right)\, \left(\frac{r}{r_0}\right)^\nu P_\nu(\cos\zeta) \label{eq:T_h} \\
	T_2(r,\zeta) &= \sum\limits_{\nu=0}^{\infty}B_\nu\,J_\nu(r) \,P_\nu(\cos\zeta) \; . \label{eq:T_i}
	\end{align}
\end{subequations}
Then, $T_1+T_2$ yield \eqref{eq:newApproximation_Tsurface} on the surface.
The particular solution employs the asymmetry factor $J_\nu$
\begin{subequations}
	\begin{align}
	J_\nu(r) &= \frac1{r_0} \left[
	r^{-\nu-1} \int\limits_{0}^{r} s^{\nu+2} q_\nu(s) \dd s + 
	r^\nu \int\limits_{r}^{r_0} s^{\nu-1} q_\nu(s) \dd s 
	\right] \\
	q_\nu(r) &= \frac{2 \nu+1}{2} \int\limits_{-1}^{1}q(r,x)\,P_\nu(x)\dd x \label{eq:q_Legendre_expansion_coefficients} \\
	J_\nu &\equiv J_\nu(r_0) = \int\limits_{0}^{r_0} \left(\frac{r}{r_0}\right)^{\nu+2} q_\nu(r) \dd r \; . \label{eq:asymmetry_factor_J}
	\end{align}
	\label{eq:asymmetry_factor_all}
\end{subequations}
$q_\nu(r)$ are the Legendre expansion coefficients of the source $q(r,\zeta)$.
For perfectly absorbing spheres, the asymmetry factors yield $J_0=1/4$ and $J_1=\pm 1/2$ (positive for irradiation into direction $\mathbf{e}_I=-\mathbf{e}_z$).

\subsubsection{Boundary conditions}\label{sec:ht_bc}
To account for thermal radiation and a temperature jump at the surface, the following boundary conditions were chosen
\begin{subequations}
	\begin{align}
		k\diffp{T}{{\mathbf{n}}} &= k_\gas\diffp{T_\gas}{{\mathbf{n}}} - \sigma_\text{SB}\varepsilon\left(T^4-T_\rad^4\right) \qquad\text{at }\partial V \label{eq:boundary_condition_HT}\\
		T_\gas - T &= \kappa_\text{t}\,r_0\,\Kn\diffp{T_\gas}{{\mathbf{n}}} \qquad\text{at }\partial V \label{eq:boundary_condition2_HT} \\
		T_\gas &\xrightarrow{r\rightarrow\infty} T_\infty \, .\label{eq:temperature_jump_condition}
	\end{align}
	\label{eq:boundary_conditions}
\end{subequations}
The last boundary condition is already met by the ansatz for the gas temperature in \eqref{eq:T_g}.
To prevent nonlinear mixing of the expansion coefficients $A_\nu$ and $B_\nu$ at multiple orders in the first boundary condition, the term $\sigma_\text{SB}\varepsilon(T^4-T_\rad^4)$ will be linearized at the mean temperature $\tilde{T}$
\begin{align}
	\sigma_\text{SB}\varepsilon(T^4-T_\rad^4) &= \sigma_\text{SB}\varepsilon\left(4 T\, \tilde{T}^3-T_\rad^4 -3 \tilde{T}^4\right) + \dots \\
	\tilde{T} &= \left( \frac1{4\pi}\int\limits_0^{2\pi}\int\limits_0^\pi T(\zeta)^4 \sin\zeta \dd\zeta\dd\xi \right)^{1/4}\; , \label{eq:mean_temperature4}
\end{align}
which is given by integrating the boundary conditions.
The second boundary condition is the temperature jump condition at the gas-particle surface. For $\Kn\to 0$ (\textit{co} regime) the sphere and the gas layer surrounding it are in thermal equilibrium. The thermal accommodation coefficient $\alpha$ defines the temperature jump coefficient $\kappa_\text{t}$ as \cite{Reed1977}
\begin{equation}
   \kappa_\text{t}(\alpha)\simeq\frac{15}{8}\left(\frac{1-\alpha}{\alpha}\right) \; . \label{eq:temperature_jump_coeff}
\end{equation}

\subsubsection{Solution}\label{sec:solution}
In a similar procedure as in Paper 1 the coefficients $A$ and $C$ can be obtained from the boundary conditions in Eqs. \ref{eq:boundary_condition_HT} and \ref{eq:boundary_condition2_HT} by using the orthogonality relations of the Legendre polynomials $P_\nu=P_\nu^0$ (\eqref{eq:orthogonality}), $B$ is obtained by the inhomogeneous heat transfer equation (\eqref{eq:heatEQ1}; upper index sf means slip flow)
\begin{subequations}
\begin{align}
   A_\nu^\sf &= \frac{I\,J_\nu}{\nu\frac{k}{r_0} + \frac{k_\gas}{r_0}\frac{\nu+1}{1+(\nu+1)\kappa_\text{t}\Kn} + 4\sigma_\text{SB}\varepsilon\,\left(\tilde{T}^\sf\right)^3} \qquad \nu\ge 1 \label{eq:A_sf} \\
   A_0^\sf &= \frac{I\,J_0 + \frac{1}{1+\kappa_\text{t}\Kn}\,\frac{k_\gas}{r_0}\,T_\infty + \sigma_\text{SB}\varepsilon\left(3\,\left(\tilde{T}^\sf\right)^4+T_\rad^4\right) }{\frac{k_\gas}{r_0}\frac1{1+\kappa_\text{t}\Kn} + 4\sigma_\text{SB}\varepsilon\,\left(\tilde{T}^\sf\right)^3} \label{eq:A0_sf} \stackrel{\eqref{eq:mean_temperature}}{=} \overline{T} \\
   B_\nu^\sf &= \frac{I \,r_0}{(2\nu+1)k} \label{eq:B_sf} \\
   C_\nu^\sf &= \frac1{1+(\nu+1)\kappa_\text{t}\Kn} A_\nu^\sf \qquad \nu\ge 1 \label{eq:C_sf} \\
   C_0^\sf &= \frac1{1+\kappa_\text{t}\Kn}\frac{I\,J_0 - 4 \sigma_\text{SB}\varepsilon\,\left(\tilde{T}^\sf\right)^3 \, T_\infty + \sigma_\text{SB}\varepsilon\left(3\,\left(\tilde{T}^\sf\right)^4+T_\rad^4\right) }{\frac{k_\gas}{r_0}\frac1{1+\kappa_\text{t}\Kn} + 4\sigma_\text{SB}\varepsilon\,\left(\tilde{T}^\sf\right)^3} \; . \label{eq:C0_sf}
\end{align}
\end{subequations}


\subsubsection{Mean temperatures}
The mean surface temperature $\overline{T}$ is solely determined by the 0-th expansion coefficient (using \eqref{eq:orthogonality})
\begin{equation}
\overline{T} = \frac1{4\pi}\int\limits_0^{2\pi}\int\limits_0^\pi T(r_0,\zeta) \sin\zeta \dd\zeta\dd\xi \stackrel{\eqref{eq:newApproximation_Tsurface}}{=} A_0 \; . \label{eq:mean_temperature} 
\end{equation}
For the gas, the mean temperature across a spherical layer is given by
\begin{equation}
\overline{T_\gas(r)} = \frac1{4\pi}\int\limits_0^{2\pi}\int\limits_0^\pi T_\gas(r,\zeta) \sin\zeta \dd\zeta\dd\xi \stackrel{\eqref{eq:T_g}}{=} T_\infty + C_0^\sf\frac{r_0}{r} \; , \label{eq:mean_temperature_gas_ball}
\end{equation}
and therefore the mean gas temperature around the particle is 
\begin{equation}
\overline{\left.T_\gas\right|_{\partial V}} = \overline{T_\gas(r_0)} = T_\infty + C_0^\sf \; . \label{eq:mean_temperature_gas_KnudsenLayer}
\end{equation}

$\tilde{T}$ can be obtained by integrating the boundary condition \eqref{eq:boundary_condition_HT} around the sphere, and using Gauss's theorem
\begin{align}
	-k\int\limits_{\partial V} \boldsymbol{\nabla}T\cdot\mathrm{d}\mathbf{A} &= \int\limits_{\partial V} \left( k_\gas\,\mathbf{n}\cdot\boldsymbol{\nabla}T + \sigma_\text{SB}\varepsilon\left(T^4-T_\rad^4\right) \right)\dd A \nonumber \\
	&= -k\int\limits_V \Delta T \dd V \nonumber \\
	&\stackrel{\text{\eqref{eq:heatEQ}}}{=} \varepsilon\, I_0 \int\limits_V q(r,\zeta)\dd V = \pi r_0^2 \,\varepsilon\, I_0 \; .
\end{align}
Then, the temperature $\tilde{T}$ (\eqref{eq:mean_temperature4}) meets the balance
\begin{equation}
\pi r_0^2\, \varepsilon \, I_0 = 4\pi r_0^2 \sigma_\text{SB}\varepsilon\left(\tilde{T}^4-T_\rad^4 \right) \; , \label{eq:mean_temperatures_balance}
\end{equation}
and is
\begin{equation}
	\tilde{T}^\sf = T_\bb := \sqrt[4]{\frac{I_0}{4\sigma_{\text{SB}}}+T_\rad^4} \label{eq:blackBodyTemp} \; .
\end{equation}
Inserted into \eqref{eq:A_sf} ($A_1^\sf$), the photophoretic force $F^\sf=F^\sf(A_1^\sf)$ will become slightly non-linear in the radiative flux $I_0$.

\subsection{Result}
Summarizing all previous findings, the photophoretic force in the slip flow regime with a gas temperature $T_\infty$ and a radiation field temperature $T_\rad$ is given as
\begin{subequations}
\begin{align}
\mathbf{F}_\text{phot}^\sf &= -4\pi\frac{\eta^2_\text{dyn} }{ \rho\,\overline{\left.T_\gas\right|_{\partial V}} } \frac{\kappa_\text{s}+3 \kappa_\text{h} \kappa_\text{m} \Kn}{1+3 \kappa_\text{m} \Kn} \cdot  \label{eq:Fphot_low_Knudsen}\\
&\quad \cdot \frac1{1+2\kappa_\text{t}\Kn}\frac{I\,J_1}{\frac{k}{r_0} + \frac{k_\gas}{r_0}\frac2{1+2\kappa_\text{t}\Kn} + 4\sigma_\text{SB}\varepsilon\,T_\bb^3} \, \mathbf{e}_z \nonumber \\
\overline{\left.T_\gas\right|_{\partial V}} &= T_\infty + \frac1{1+\kappa_\text{t}\Kn}\frac{I\,J_0 - 4 \sigma_\text{SB}\varepsilon\,T_\bb^3 \, T_\infty + \sigma_\text{SB}\varepsilon\left(3\,T_\bb^4+T_\rad^4\right) }{\frac{k_\gas}{r_0}\frac1{1+\kappa_\text{t}\Kn} + 4\sigma_\text{SB}\varepsilon\,T_\bb^3} \\
T_\bb &= \sqrt[4]{\frac{I_0}{4\sigma_{\text{SB}}}+T_\rad^4} \; .
\end{align}
\label{eq:Fphot_low_Knudsen_all}
\end{subequations}
Apart from the additional radiative term $4\sigma_\text{SB}\varepsilon\,T_\bb^3$, the results are in agreement with Eq. 36 from \cite{Reed1977} \footnote{\cite{Reed1977} incorporates the radiation source term into the boundary condition, with $J_1=1/2$.} and Eq. 29 from \cite{Mackowski1989aerosol} for $\kappa_\text{h}=0$.

Eventually, the phothophoretic velocity is given as (\eqref{eq:uphot})
\begin{align}
	\mathbf{u}_\text{phot}^\sf &= -\frac23\frac{\eta_\text{dyn} }{ \rho\,\overline{\left.T_\gas\right|_{\partial V}}\,r_0 } \frac{\kappa_\text{s}+3 \kappa_\text{h} \kappa_\text{m} \Kn}{1+2 \kappa_\text{m} \Kn} \cdot \label{eq:uphot_low_Knudsen}\\
	&\quad\cdot \frac1{1+2\kappa_\text{t}\Kn}\frac{I\,J_1}{\frac{k}{r_0} + \frac{k_\gas}{r_0}\frac2{1+2\kappa_\text{t}\Kn} + 4\sigma_\text{SB}\varepsilon\,T_\bb^3} \, \mathbf{e}_z \; . \nonumber
\end{align}

\subsection{Continuum limit}\label{sec:co_phothophoresis}
The rotationally symmetric solution above (\eqref{eq:Fphot_low_Knudsen_all}) is reused to describe the force in the \textit{co} limit ($\Kn\rightarrow 0$).
From the boundary condition \eqref{eq:temperature_jump_condition} it is $\overline{\left.T_\gas\right|_{\partial V}} = \overline{T}$ at the surface, and therefore the force reads 
\begin{subequations}
\begin{align}
	\mathbf{F}_\text{phot}^\co &\stackrel{\text{\eqref{eq:Fphot_low_Knudsen}}}{=} -4\pi\,\kappa_\text{s}\,\frac{\eta^2_\text{dyn}}{\rho \, \overline{T}} \, \frac{I\,J_1}{\frac{k}{r_0} + 2\frac{k_\gas}{r_0} + 4\sigma_\text{SB}\varepsilon\,T_\bb^3} \, \mathbf{e}_z \label{eq:Fphot_co} \\
	\overline{T} &\stackrel{\text{\eqref{eq:mean_temperature_gas_KnudsenLayer}}}{=} T_\infty +  \lim\limits_{\Kn\to 0} C_0^\sf  \\ &= T_\infty + \frac{I\,J_0 - 4 \sigma_\text{SB}\varepsilon\,T_\bb^3 \, T_\infty + \sigma_\text{SB}\varepsilon\left(3\,T_\bb^4+T_\rad^4\right) }{\frac{k_\gas}{r_0} + 4\sigma_\text{SB}\varepsilon\,T_\bb^3} \nonumber \\
	&\stackrel{\text{\eqref{eq:mean_temperature}}}{=} \lim\limits_{\Kn\to 0} A_0^\sf = \frac{I\,J_0 + \frac{k_\gas}{r_0}\,T_\infty + \sigma_\text{SB}\varepsilon\left(3\,T_\bb^4+T_\rad^4\right)}{\frac{k_\gas}{r_0} + 4\sigma_\text{SB}\varepsilon\,T_\bb^3} \; .  \label{eq:A0_co}
\end{align}
 \label{eq:Fphot_co_all}
\end{subequations}
The photophoretic velocity is
\begin{align}
   \mathbf{u}_\text{phot}^\co &\stackrel{\text{\eqref{eq:uphot_low_Knudsen}}}{=} -\frac23\,\kappa_\text{s}\,\frac{\eta_\text{dyn} }{ \rho\,\overline{T}\,r_0 } \, \frac{I\,J_1}{\frac{k}{r_0} + 2\frac{k_\gas}{r_0} + 4\sigma_\text{SB}\varepsilon\,T_\bb^3} \, \mathbf{e}_z \; . \label{eq:uphot_co}
\end{align}
\section{INTERPOLATING BETWEEN \textit{fm}- AND \textit{co}-PHOTOPHORESIS}\label{sec:tr_phothophoresis}
\begin{figure}[!h]
	\centering
	\includegraphics[width=\columnwidth]{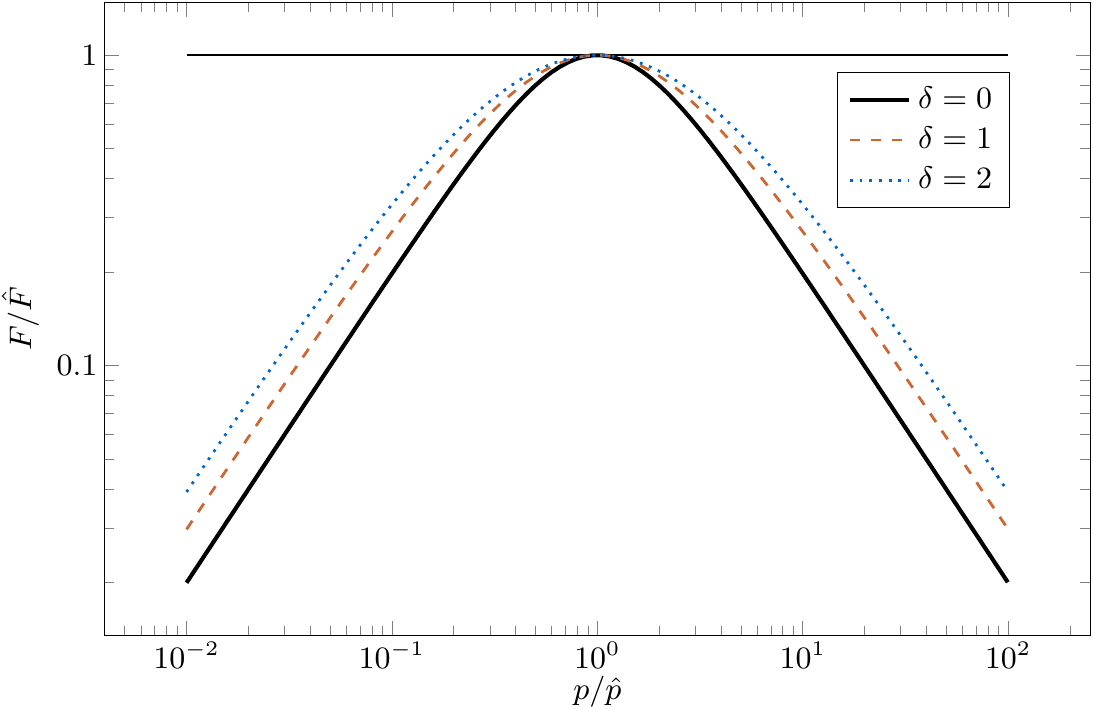}
	\caption{\label{fig:fPhot-fPhotMax}Interpolation between \textit{fm} and \textit{co} regimes. The photophoretic force peaks at $\hat{F}=F(\hat{p})$.}
\end{figure}

An empirical method is used to describe the photophoretic force in the transition regime due to the complexity of transport processes in this regime.
\cite{Rohatschek1995} presents a phenomenological equation satisfying the linear proportionality of the force with the pressure in the \textit{fm} regime and the inverse proportionality in the \textit{co} regime by
\begin{equation}
   \frac{F_\text{phot}}{\hat{F}} = \frac{2+\delta}{\frac{p}{\hat{p}}+\delta+\frac{\hat{p}}{p}} \; , \label{eq:Fphot_Transition}
\end{equation}
with the free parameter $\delta$ to be adjusted along the experimental values (\fig{fig:fPhot-fPhotMax}).
Because of the changing proportionality at an unknown pressure $\hat{p}$, the force peaks at $\hat{F}=F(\hat{p})$.
\cite{Hettner1928} already suggests the same equation with $\delta=0$. \cite{Rohatschek1995} also favors it, justifying it to be the best-fitting version of conducted experiments in the past, including work of other researchers such as \cite{Tong1975,Arnold1980}. Nonetheless, the experiments of \cite{Rosen1964} with large carbon agglomerates do not obey above's law. \cite{Rohatschek1985Exp} gave evidence that for large agglomerates, theories of $\Delta T$-photophoresis cannot be applied because of the superposition of $\Delta\alpha$- and $\Delta T$-photophoresis. One of the experimental results mentioned by \cite{Rohatschek1995} implied $\delta = 0.8$.
The gas-kinetic calculations made by \cite{Chernyak1993} suggested $\delta\simeq 1$, and for slip-flow theories, e.g. in \cite{Reed1977} it is even $\delta\ge 2$, both fitting about 67\% and less than 50\%, respectively, of the experimental findings \cite{Rohatschek1995} discussed.

\cite{Hettner1928} suggested an interpolation (Eq. 20 in the respective publication), which is
\begin{equation}
   \frac1{F_\text{phot}} = \frac1{F^\text{co}_\text{phot}} + \frac1{F^\text{fm}_\text{phot}} \; . \label{eq:Fphot_Transition2}
\end{equation}
We therefore present a new interpolation along \cite{Rohatschek1995} for $\delta=0$, whose scope of validity includes not only $|\overline{T}/T_\infty|\simeq 1$ but also stronger temperature deviations and therefore higher intensities.
This interpolation is based on the \textit{fm} equation from Paper 1 and the \textit{co} equation from this paper (\eqref{eq:Fphot_co_all}).

\subsection{Longitudinal photophoresis in the transition regime}\label{sec:longitudinal_photophoresis_in_the_transition_regime}
To determine $\hat{F}$ and $\hat{p}$ for longitudinal photophoresis in the description of \eqref{eq:Fphot_Transition} for $\delta=0$, a few more steps have to be made.
Starting with \eqref{eq:Fphot_Transition2}, the force in the \textit{fm} and the \textit{co} regimes is ($A_0^\co := \lim\limits_{\Kn\to 0}A_0^\sf$ etc., $A_\nu^\co \equiv C_\nu^\co$, $A_1^\fm = \frac{I\,J_1}{\frac{k}{r_0} + h + 4\sigma_\text{SB}\varepsilon\,\left(\tilde{T}^\fm\right)^3}$ (Paper 1), and changing the notation used in Paper 1 from $T_\gas^\-$ to $T_\infty$)
\begin{subequations}
\begin{align}
   F^\text{fm}_\text{phot} &\stackrel{\text{Paper 1}}{\simeq} \frac{\pi}{3} \, \alpha \, \alpha_\text{m} \, \frac{p}{\sqrt{T_\infty\overline{T_\gas^\+}}} \, r_0^2 \, A_1^\fm \nonumber \\
   &\simeq 2\,\Xi\,\frac{p}{p^*}\,\tau^\text{fm} \,r_0\,J_1\, I \label{eq:Fphot_Xi_fm} \\
   F^\text{co}_\text{phot} &\stackrel{\text{\eqref{eq:Fphot_co}}}{=} 4\pi\,\kappa_\text{s}\,\frac{\eta^2_\text{dyn}}{\rho \, A_0^\co}\, A_1^\co \nonumber \\
   &= 2\,\Xi\,\frac{p^*}{p}\,\tau^\text{co}\,r_0\,J_1\, I \label{eq:Fphot_Xi_co} \; ,
\end{align}
\end{subequations}
where the mean scattered gas temperature $\overline{T_\gas^\+}$, the constant $\Xi$ and the characteristic pressure $p^*$ are (like in \cite{Rohatschek1995})
\begin{subequations}
	\begin{align}
	\overline{T_\gas^\+} &= T_\infty+\alpha\left(A_0^\fm-T_\infty\right) \nonumber &\text{\eqref{eq:mean_T_plus}}\\
	A^\fm_0 &\stackrel{\text{Paper 1}}{=} \frac{I\,J_0 + h\,T_\infty + \sigma_\text{SB}\varepsilon\,\left(3(\tilde{T}^\fm)^4+T_\rad^4\right)}{h + 4\sigma_\text{SB}\varepsilon\,\left(\tilde{T}^\fm\right)^3} \label{eq:A0_fm} \\
	h &\stackrel{\text{Paper 1}}{=} \frac{1}{2} \alpha_{\rm m} \alpha \frac{p}{T_\infty} v_\text{th} \label{eq:h_fm} \\
	\Xi &= \frac{\pi}{2}\,\sqrt{\frac{\pi}{3}}\,\frac{v_\text{th}\,\eta_\text{dyn}}{\sqrt{A_0^\co\sqrt{T_\infty\overline{T_\gas^\+}}}}\label{eq:Xi}\\
	p^* &= \frac12\sqrt{3\pi}\,\frac{v_\text{th}\,\eta_\text{dyn}}{r_0} = \frac{3}{\pi} \, \Xi \, \frac{\sqrt{A_0^\co\sqrt{T_\infty\overline{T_\gas^\+}}}}{r_0} \label{eq:char_p}\\
	A_0^\co &= \frac{I\,J_0 + \frac{k_\gas}{r_0}\,T_\infty + \sigma_\text{SB}\varepsilon\left(3\,T_\bb^4+T_\rad^4\right)}{\frac{k_\gas}{r_0} + 4\sigma_\text{SB}\varepsilon\,T_\bb^3} \; . \nonumber &\text{\eqref{eq:A0_co}}
	\end{align}
	\label{eq:interpolationvalues}
\end{subequations}
Here, the ideal gas equation $p = \frac{\rho}{M}R_\gas\,T_\gas$ was used to express the mean thermal gas speed as $v_\text{th} = \sqrt{8p/(\pi\rho)}$. \eqref{eq:h_fm} is valid for mono-atomic gas, for di-atomic gas the factor $1/2$ has to be replaced with $3/4$ \citep{Rohatschek1985}.
The mean temperatures $\tilde{T}^\fm$ and $\overline{T}$ for $h>0$ can be determined by solving \eqref{eq:A0_fm} iteratively with $\tilde{T}^\fm=\overline{T}$ starting at $T_\bb$. For a relatively small $h$, $\tilde{T}^\fm = T_\bb$ (Paper 1).
The dimensionless scaling coefficients $\tau$ are subsequently
\begin{subequations}
	\begin{align}
		\tau^\text{fm} &= \frac{\sqrt{A_0^\co}}{\sqrt[4]{T_\infty\overline{T_\gas^\+}}} \, \frac{\alpha\,\alpha_\text{m}}{2} \, \frac{1}{\frac{k}{r_0}+h+ 4\sigma_\text{SB}\varepsilon\,\left(\tilde{T}^\fm\right)^3 }\\
		\tau^\text{co} &= \frac{\sqrt[4]{T_\infty\overline{T_\gas^\+}}}{\sqrt{A_0^\co}}\,\kappa_\text{s} \, \frac{1}{\frac{k}{r_0}+2\frac{k_\gas}{r_0}+ 4\sigma_\text{SB}\varepsilon\,T_\bb^3 } \; .
	\end{align}
	\label{eq:taus}
\end{subequations}

The interpolation equation \eqref{eq:Fphot_Transition2} enables --- together with the equations above --- to derive
\begin{subequations}
\begin{align}
   \hat{F} &= \Xi\,\sqrt{\tau^\text{co}\,\tau^\text{fm}}\,r_0 \, J_1\, I \label{eq:F_max} \\
   \hat{p} &= \sqrt{\frac{\tau^\text{co}}{\tau^\text{fm}}}\,p^* \label{eq:p_max} \; .
\end{align}
\end{subequations}
Compared to the work in \cite{Rohatschek1995}, the maximum force $\hat{F}$ is determined by the geometric mean of $\tau^\text{co}$ and $\tau^\text{fm}$ as additional factor.
Similarly, the pressure $\hat{p}$ where the forces maximizes is given by extending the result in \cite{Rohatschek1995} by an additional factor, i.e. the square root of the ratio of the two $\tau$.
\section{DISCUSSION}\label{sec:discussion}
In this section, the underlying \textit{fm} and \textit{co} limit approximations are discussed.
A brief comparison to the original model by \cite{Rohatschek1995,Hettner1928} is given afterwards.

\subsection{\textit{fm} limit equation accuracy}
In Paper 1 a new approximation for photophoretic forces in the \textit{fm} regime following
\begin{equation}
	F^\text{fm}_\text{phot} \simeq \frac{\pi}{3} \, \alpha \, \alpha_\text{m} \, \frac{p}{\sqrt{T_\infty\overline{T_\gas^\+}}} \, r_0^2 \, \frac{I\,J_1}{\frac{k}{r_0} + h + 4\sigma_\text{SB}\varepsilon\,T_\bb^3}
	\label{eq:paper1}
\end{equation}
was introduced.
As shown in that paper, this formula for photophoresis on spherical particles with surface temperatures strongly deviating from the gas temperature $T_\infty$ or high intensities $I$ significantly increases the accuracy of analytically determined photophoretic forces with respect to numerical values.
Different classic approximation for the photophoretic force in the \textit{fm} regime which are not supporting these gas temperatures and intensity conditions were compared to the new approximation to emphasize the need for an additional equation in the \textit{fm} regime with an extended scope of validity.
While still covering the classic scope of validity (for $\alpha_\text{m}=1$, this equation can very well be approximated by the \textit{fm} equation from \cite{Beresnev1993} for $\alpha_\textbf{n}=1$), \eqref{eq:paper1} has an average relative error of about 1\% for particles with a radius of up to $1.1\u{mm}$. With a maximum and minimum relative errors of only 7\% and $\approx\!-50\%$, respectively, (for details see Paper 1), it is far more reliable under rather extreme conditions than the classic \textit{fm} approximations, which then overestimate the force up to orders of magnitude, as they were designed for basically low intensities.

\subsection{\textit{co} limit equation accuracy}
\begin{table}[h]
	\centering
	\caption{\label{tab:parameter_config}
		Intervals for the parameter sweep in COMSOL, where the heat transfer equation (\eqref{eq:heatEQ1}) with the boundary condition given by \eqref{eq:boundary_condition_HT} was solved ($[a,b]$ denotes an interval between the numbers $a$ and $b$).
		All intervals are equally subdivided (log scale; the additional `$1\u{m}$' for $r_0$ means, there is a gap between $1\u{m}$ and $0.11\u{m}$ concerning this equal subdivision).		
		Details on the subdivision can be found in \cite{ChrisDiss}.
		}
	\begin{tabular}{c L{5.5cm}}
		\toprule
		parameter & parameter sweep intervals \\ \midrule
		$r_0$ & $[1.1\times 10^{-4}, 1.1\times 10^{-1}]\u{m}$, and 1 m\\
		$k$ & $[10^{-3}, 8]\u{W\,m^{-1}\,K^{-1}}$\\
		$I$ & $[0.5, 40]\u{kW\,m^{-2}}$\\
		$T_\rad$ & $[0, 350]\u{K}$ \\ \bottomrule
	\end{tabular}
\end{table}
\begin{figure}[h!]
	\resizebox{1.0\columnwidth}{!}{\input{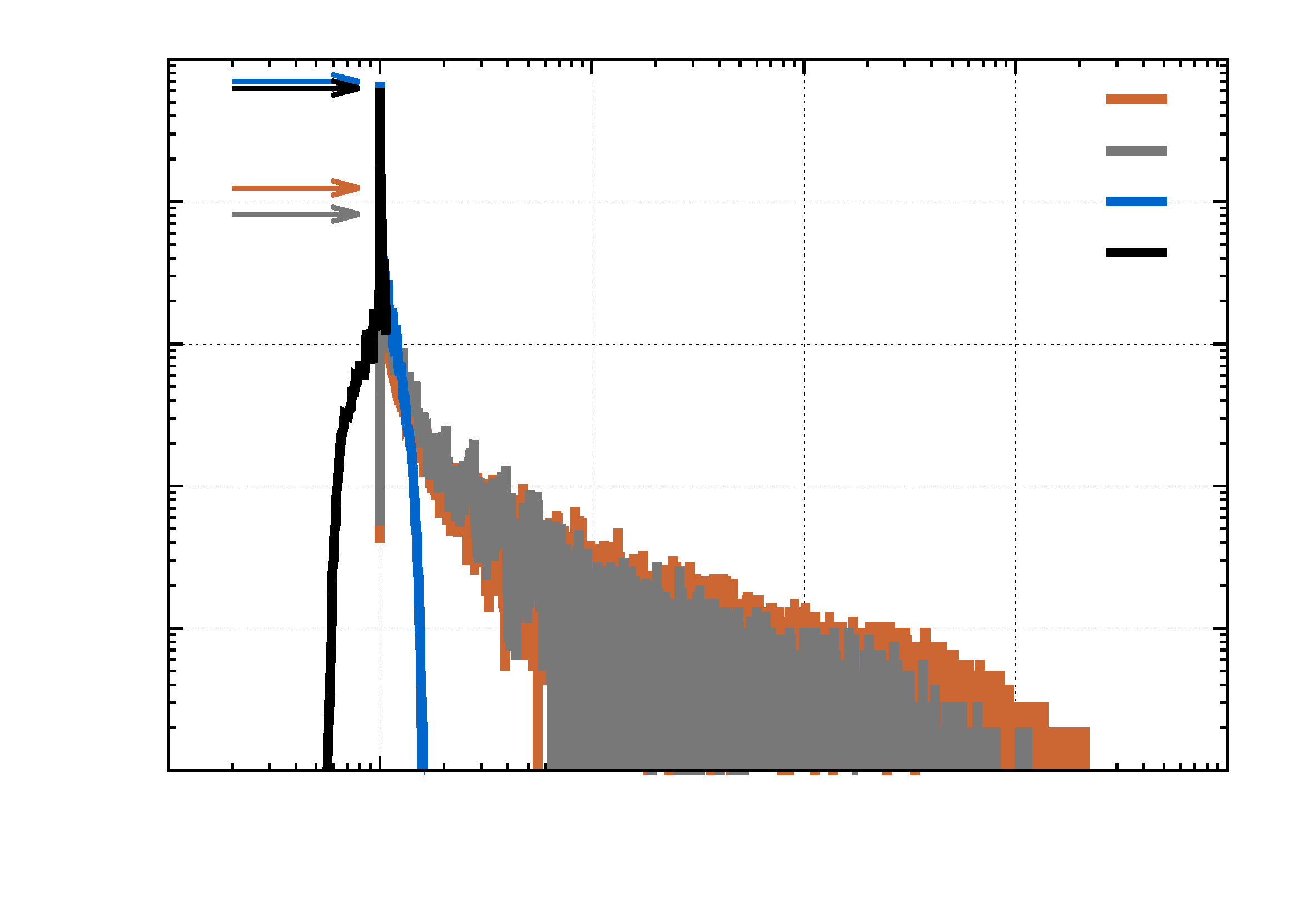}}
	\caption{
		Parameter sweep histogram for $196\,344$ parameter combinations.
		The parameter sweep intervals are given in \tab{tab:parameter_config}.
		Like in Paper 1, the effectively exact numerical result was obtained with COMSOL by solving the heat transfer equation (\eqref{eq:heatEQ1}) with the boundary condition given by \eqref{eq:boundary_condition_HT}.
		The bin size is 0.005 (0.5\%).
		Color-coded arrows point towards the respective histogram's peak.
		The histogram for $A_1^\co$ (black) is restricted to $0.57\le A_1^\co/A_1\le1.07$ while the other expansion coefficients $A_1^{\co,0}$ \citep{Reed1977,Mackowski1989aerosol} and $A_1^{\co,\rad}$ are overestimated up to several orders of magnitude.
		The particle mean temperature $A_0^\co$ (blue) is also very close to the exact value $A_0$, the ratio is within $1.00\le A_0^\co/A_0\le1.63$. 
	}
	\label{fig:quality}
\end{figure}
\begin{table*}[!htbp]
\centering
\caption{
	Statistical properties of the ratio of the particle temperature expansion coefficients (see \eqref{eq:a1}).
	A parameter sweep of $196\,344$ parameter combinations was performed along the parameter intervals given in \tab{tab:parameter_config}.
	Values in round brackets are for $r_0$ restricted to the interval $[0.11,11]\u{mm}$.
	 }
\label{tab:statistics}
\begin{tabular}{$C{5cm} >{\bfseries}^C{1cm}>{\bfseries}^C{1cm} ^C{1cm}^C{1cm}>{\itshape}^C{1cm}}
\toprule
ratio of particle temperature expansion coefficients: analytic/numerical & min & max & mean & median & STD \\
\midrule
$A_1^{\co,0}/A_1$ 	& 1.00 \textcolor{gray}{(1.00)}	& 38\,368 \textcolor{gray}{(424)} & 46.9 \textcolor{gray}{(7.23)}	& 2.25 \textcolor{gray}{(1.35)}	& 357 \textcolor{gray}{(20.1)} \\ \hline
$A_1^{\co,\rad}/A_1$	& 1.00 \textcolor{gray}{(1.00)} 	& 20\,802 \textcolor{gray}{(266)} & 8.10 \textcolor{gray}{(3.00)} 	& 1.45 \textcolor{gray}{(1.19)}	& 83.2 \textcolor{gray}{(7.78)} \\ \hline
\rowstyle{\bfseries} $A_1^{\co}/A_1$	& 0.57 \textcolor{gray}{(0.66)}	& 1.07 \textcolor{gray}{(1.07)} & 0.97 \textcolor{gray}{(0.99)} 	& 1.00 \textcolor{gray}{(1.00)}	& 0.08 \textcolor{gray}{(0.05)} 	\\ \hline
\rowstyle{\bfseries} $A_0^{\co}/A_0$	& 1.00 \textcolor{gray}{(1.00)} 	& 1.63 \textcolor{gray}{(1.43)} & 1.06 \textcolor{gray}{(1.04)} 	& 1.01 \textcolor{gray}{(1.00)}	& 0.10 \textcolor{gray}{(0.07)} \\ \hline
\rowstyle{\bfseries} $\frac{A_1^{\co}}{A_0^{\co}}/\frac{A_1}{A_0}$	& 0.35 \textcolor{gray}{(0.46)} 	& 1.05 \textcolor{gray}{(1.05)} & 0.93 \textcolor{gray}{(0.96)} 	& 1.00 \textcolor{gray}{(1.00)}	& 0.13 \textcolor{gray}{(0.10)} \\  \bottomrule
\end{tabular}
\end{table*}
\begin{figure}[h!]
	\centering
	\begin{subfigure}[t]{1\columnwidth}
		\resizebox{1.0\columnwidth}{!}{\input{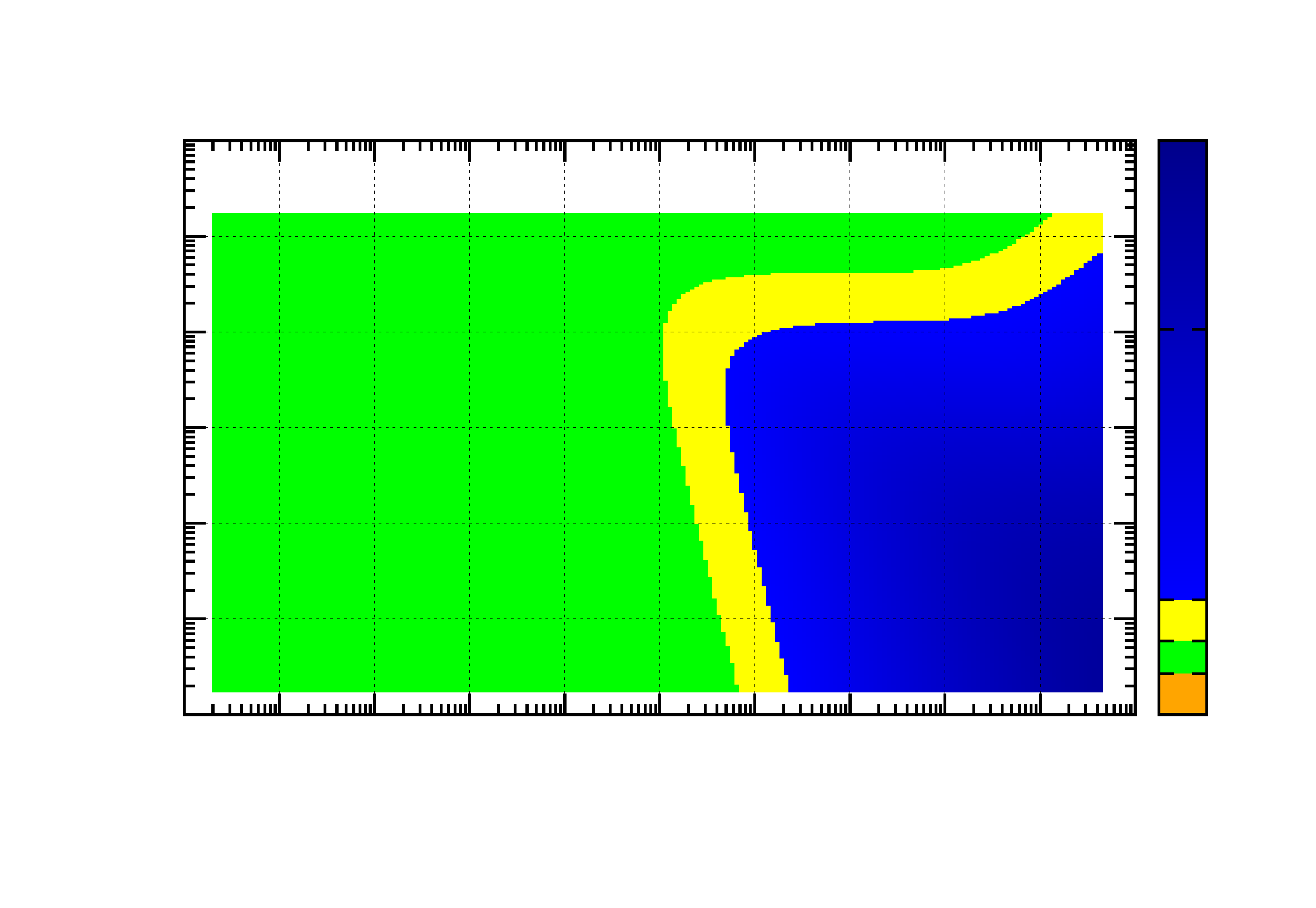}}
		\caption{$A_0^{\co}/A_0$}
	\end{subfigure}%
	
	\begin{subfigure}[t]{1\columnwidth}
		\centering
		\resizebox{1.0\columnwidth}{!}{\input{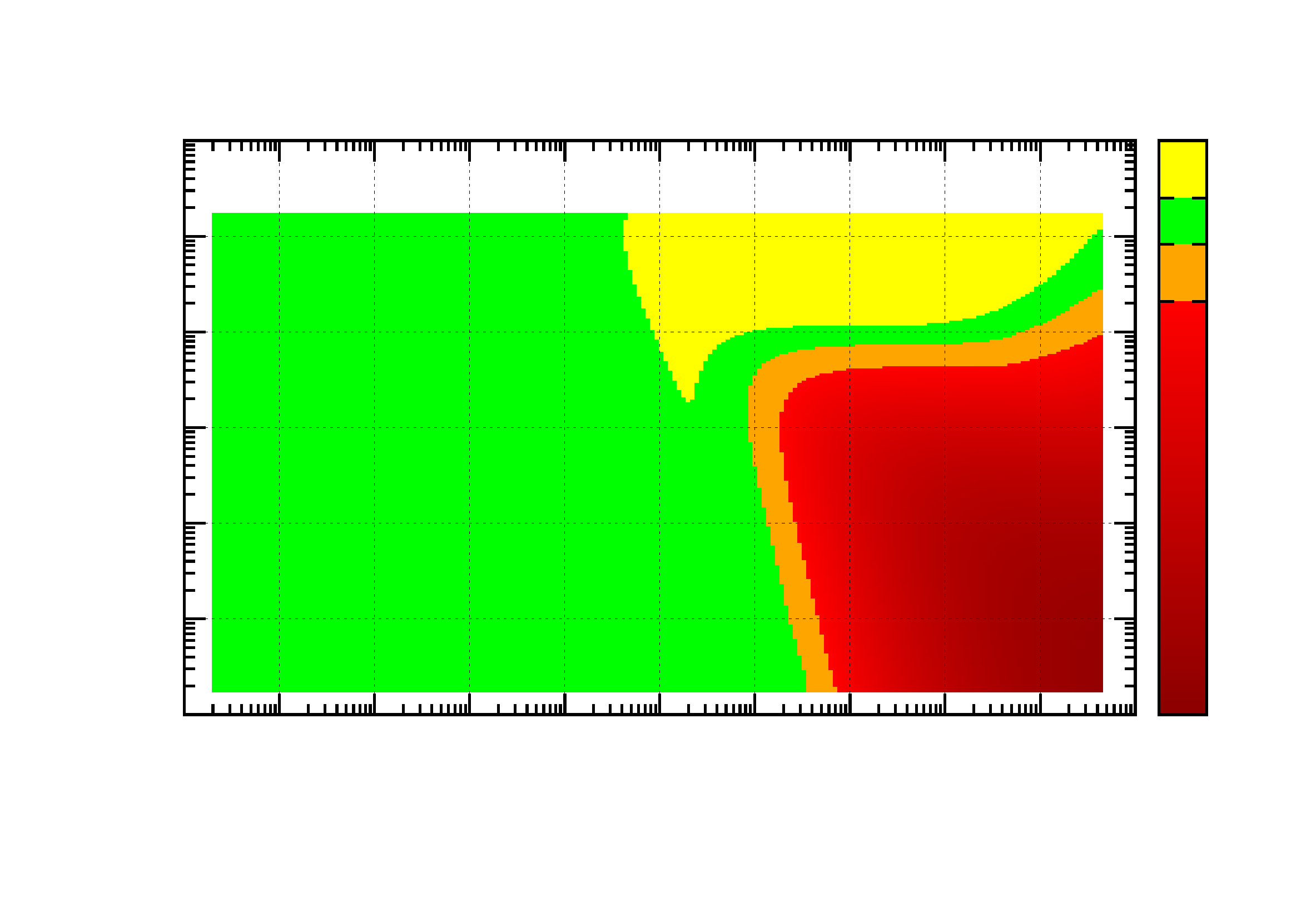}}
		\caption{\label{fig:phi_theta_A1}$A_1^{\co}/A_1$}
	\end{subfigure}
	\caption{\label{fig:phi_theta}
		Ratio of the particle temperature expansion coefficients. The dimensionless variables $\varphi_\rad$ and $\vartheta_\rad$ are defined in \eqref{eq:ht_var}.
		The plot for $\frac{A_1^{\co}}{A_0^{\co}}/\frac{A_1}{A_0}$ is very similar to \fig{fig:phi_theta_A1}, basically only varying in the bounds of the ratio: $0.35\le\frac{A_1^{\co}}{A_0^{\co}}/\frac{A_1}{A_0}\le1.05$, therefore we refrain from plotting this.
		}
\end{figure}

In this paper, we use the same radiation term in the boundary condition as in the previous paper (see Section \ref{sec:ht_bc}).
As the force depends on the first expansion coefficient of the gas temperature $C_1^\sf\propto A_1^\sf$ ($C_1^\co = A_1^\co$) very close to the surface, a more accurate expansion coefficient $A_1$ will obviously also improve the quality of the calculated force. This is especially true for high intensities $I$, where the radiation term $4\sigma_\text{SB}\varepsilon\,T_\bb^3$ will strongly contribute to the solution.
As the description of the entire pressure regimes photophoresis in this paper is based on the interpolation between the \textit{fm} and \textit{co} approximations, only the thermal radiation contributes as additional term in comparison to \cite{Rohatschek1995}, while those boundary conditions which are linear in the Knudsen-number disappear in the \textit{co} limit.
We performed a parameter sweep along the values in \tab{tab:parameter_config} and visualize the strong influence of the black body radiation term in the histogram in \fig{fig:quality}, where the histograms of each ratio of one of the three expansion coefficients \eqref{eq:a1} and its true value are shown. This true value was obtained from temperature distribution across the spheres, calculated with COMSOL.
The boundary conditions used in COMSOL are \eqref{eq:boundary_condition_HT}.
\tab{tab:statistics} shows minimum and maximum ratios. Beside that, it also shows other distribution information, which does not have a strict mathematical meaning but show a tendency, just as \fig{fig:quality}.
For simplicity, we restrict ourselves to the \textit{co} limit in this discussion and neglect $k_\gas$ in the COMSOL calculations, as for gases like air, high intensities and not too small particles the radiation term dominates $2k_\gas/r_0\ll4\sigma_\text{SB}\varepsilon\,T_\bb^3$.
In the other cases, the consideration of $k_\gas$ will not prevent any error here but only complicate our considerations, because the term $2\frac{k_\gas}{r_0}$ in the expansion coefficients did not arise from any linearizations or simplifications in the boundary condition.
To investigate the influence of the black body radiation term in the first expansion coefficient of the particle surface temperature $A_1$, we therefore either set $\tilde{T}$ to 0, $T_\infty$ and our result $T_\bb$:
\begin{subequations}
	\begin{align}
		\text{no thermal radiation:}\quad A_1^{\co,0} &= \frac{I\,J_1}{\frac{k}{r_0} + 2\frac{k_\gas}{r_0}} \label{eq:a1_noRad}\\
		\text{simple thermal rad.:}\quad A_1^{\co,\rad} &= \frac{I\,J_1}{\frac{k}{r_0} + 2\frac{k_\gas}{r_0} + 4\sigma_\text{SB}\varepsilon\,T_\rad^3}\label{eq:a1_simpleRad} \\
		\text{black body rad.:}\quad A_1^{\co,\bb} &= \frac{I\,J_1}{\frac{k}{r_0} + 2\frac{k_\gas}{r_0} + 4\sigma_\text{SB}\varepsilon\,T_\bb^3} \label{eq:a1_bb} \; .
	\end{align}
	\label{eq:a1}
\end{subequations}
The first equation $A_1^{\co,0}$ was obtained by \cite{Reed1977,Mackowski1989aerosol} as they did not include thermal radiation \footnote{\cite{Reed1977}: $J_1=1/2$ here}.
The second equation resembles the term used in the \textit{fm} approximation by \cite{Beresnev1993}, and the last equation is our previously obtained result (see \eqref{eq:A_sf}).
\fig{fig:quality} clearly shows the good performance of $A_1^{\co,\bb}$, i.e. when the black body temperature is used.
Surprisingly, the coefficient with no radiation $A_1^{\co,0}$ and the one that assumes the particle to radiate with $T_\rad$ both perform equally bad, although $A_1^{\co,0}$ with $k_\gas=0$ belongs to a boundary condition that does not allow a steady state solution of the heat transfer equation.
As in the \textit{co} limit, the photophoretic force depends on $1/\overline{T}$, the histogram of the ratio of the mean temperature $\overline{T}=A_0^\co$ and its numerically obtained value is also shown. It is mirrored at the ratio 1.
In Paper 1 it was shown, that these two dimensionless variables
\begin{subequations}
\begin{align}
	\varphi_\rad &= \dfrac{\varepsilon\,I_0\,r_0}{k\,T_\rad} \label{eq:phi} \\
	\vartheta_\rad &= \sigma_\text{SB}\frac{T_\rad^4}{I_0} \label{eq:theta} \; .
\end{align}
\label{eq:ht_var}
\end{subequations}
characterize different results of the heat transfer problem (scaled to unit sphere; here for omitted $k_\gas$). In \fig{fig:phi_theta} the ratio of $A_0^\co$ and $A_1^\co$ to their respective exact numerical values are plotted over $\varphi_\rad$ and $\vartheta_\rad$.
From the plots one can conclude, that in the given parameter range (see \tab{tab:parameter_config}),
the relative error of $A_0^\co$ and $A_1^\co$ is less than $2\%$ for $\varphi_\rad < 1$.
Within the model, the results in \eqref{eq:Fphot_co_all} carry about the same error.

\FloatBarrier
\subsection{Changes for the entire range of pressures}
As the \textit{fm} approximation shows even smaller relative errors for the same parameter sweep (Paper 1), the interpolation is based on two robust equations, that are the \textit{fm} and \textit{co} limit approximation of the photophoretic force.

\begin{figure}[!h]
	\centering
	\includegraphics[width=\columnwidth, trim= 3 0 38 0]{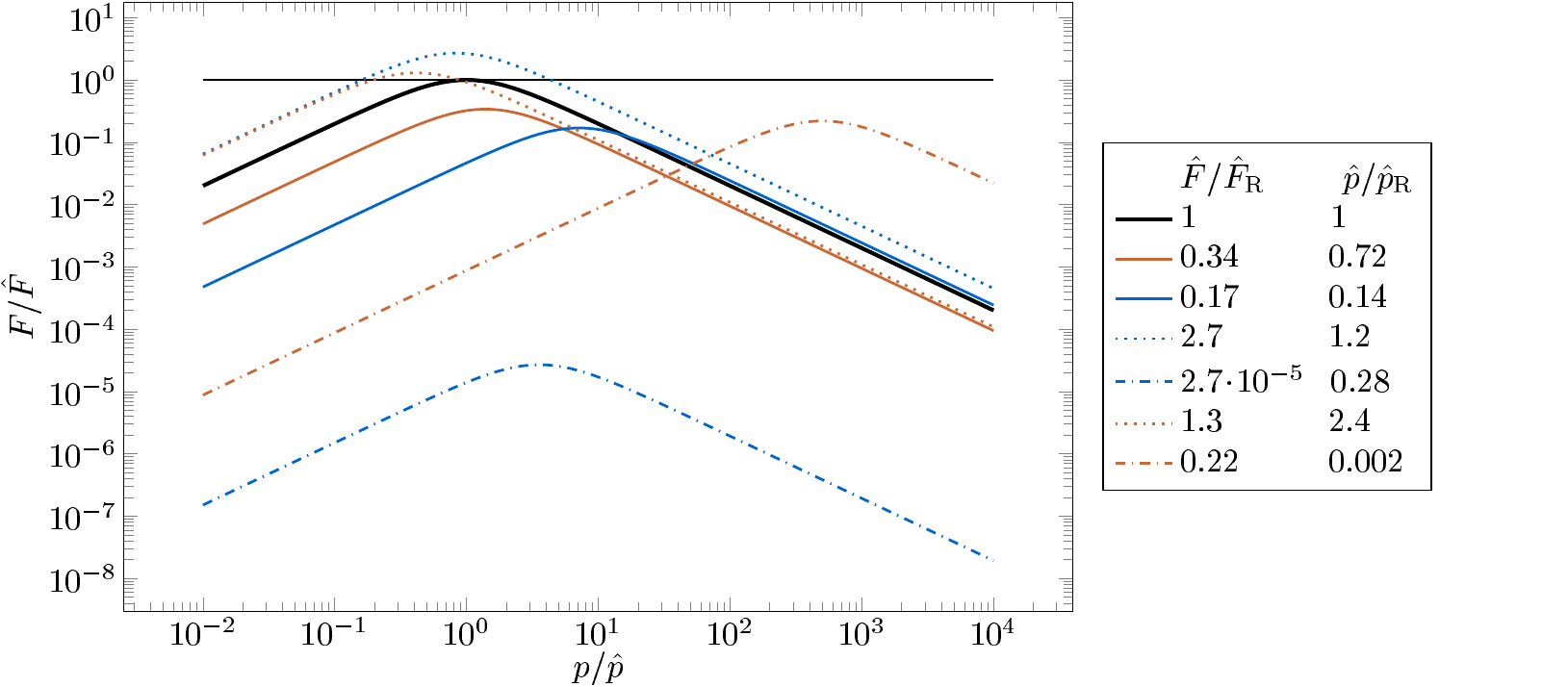}
	\caption{\label{fig:fPhot-fPhotMax2} Two case studies as well as the maximum alteration of the interpolation model $F_\text{phot}=F_\text{phot}(\hat{F},\hat{p})$ with respect to the model in \cite{Rohatschek1995} along \tab{tab:stats_interpolation}.}
\end{figure}

In the following we discuss the predictions of this model for the entire range of pressures and compare them to those made in \cite{Rohatschek1995}:
\begin{subequations}
	\begin{align}
		\hat{F}_\text{R} &= \Xi_\text{R}\,\sqrt{\kappa_\text{s}\frac{\alpha}{2}}\frac{r_0}{k}\,r_0 \, J_1\, I \label{eq:characteristic_force_R} \\
		\hat{p}_\text{R} &= \sqrt{\kappa_\text{s}\frac{2}{\alpha}} \, p^* \label{eq:characteristic_pressure_R} \\
		\Xi_\text{R} &= \frac{\pi}{2}\,\sqrt{\frac{\pi}{3}}\,\frac{v_\text{th}\,\eta_\text{dyn}}{T_\infty} \\
		p^* &= \frac12\sqrt{3\pi}\,\frac{v_\text{th}\,\eta_\text{dyn}}{r_0} \; .
	\end{align}
\end{subequations}
Just as the underlying \textit{fm} and \textit{co} approximations for the interpolation model in \cite{Rohatschek1995}, the model is also assuming very low deviances from gas and mean surface temperature.
Additionally, for simplicity, \cite{Rohatschek1995} omitted $k_\gas$ in $A_1^{\co,0} = \frac{I\,J_1}{\frac{k}{r_0} + 2\frac{k_\gas}{r_0}}$ (\eqref{eq:a1_noRad}) in his model so that $A_1^{\co,0}$ is equal to $A_1^{\fm,0} = r_0\frac{I\,J_1}{k}$ ($h=0$, too).
For very low gas heat conductivities $k_\gas$ such as air this will not introduce a significant error, but for hydrogen-helium gases it will. Beside that, the introduced error will grow strongly as the discussed particles get larger (see Paper 1).
In our model this is not the case anymore.
But for low intensities $I$ and low gas heat conductivity $k_\gas$, the interpolation proposed in this paper is basically the same as in \cite{Rohatschek1995}, which performs well (see \cite{Rohatschek1995} and Sec. \ref{sec:tr_phothophoresis}) within its scope.
We calculated the changes of $\hat{F}$ and $\hat{p}$ with respect to the values obtained with the model from \cite{Rohatschek1995} ($\hat{F}_\text{R}$ and $\hat{p}_\text{R}$).
For extreme values, the force ratio $\hat{F}/\hat{F}_\text{R}$ can reach values between $2.7\!\cdot\!10^{-5}$ and $2.7$.
The minimum pressure ratio $\hat{p}/\hat{p}_\text{R}$ can be as low as $0.13$, the maximum one $1.8$.
\fig{fig:fPhot-fPhotMax2} shows photophoretic forces for these extreme values as well as two more realistic cases in comparison to the predictions made in \cite{Rohatschek1995}.
The corresponding parameters and values are listed in \tab{tab:stats_interpolation}.
We chose a laser illuminated mm-sized particle in a cooled experimental setup and a particle in an astrophysical context as example studies which results in force/pressure ratios of $\hat{F}/\hat{F}_\text{R} = 0.34$ with $\hat{p}/\hat{p}_\text{R} = 0.72$, and $\hat{F}/\hat{F}_\text{R} = 0.17$ with $\hat{p}/\hat{p}_\text{R} = 0.15$, respectively. Both values --- especially in the last case --- show significantly different predictions.
However, experimental investigations on the interpolation for high intensities are subject to future work at this moment and beyond the scope of this paper.
One should mention here that rotation of illuminated particles --- which are observed especially in experimental studies e.g. by \cite{vanEymeren2012} --- have high influence on the photophoretic force \citep{Loesche2014}.
\begin{table*}[!h]
	\centering
	\caption{\label{tab:stats_interpolation}
		Changes of $\hat{F}$ and $\hat{p}$ with respect to the values obtained with the model from \cite{Rohatschek1995}.
		Extreme situations as well as two example studies are also sketched in \fig{fig:fPhot-fPhotMax2}.
		}
	\begin{tabular}{l l l l l l l l l}
		\toprule
							&		$I$ 				& 	$k$					&	$r_0$			&	$k_\gas$			&	$T_\infty$			&	$T_\rad$		&  	$\hat{F}/\hat{F}_\text{R}$	&  	$\hat{p}/\hat{p}_\text{R}$	\\ 
		\qquad\qquad in					&		\footnotesize{$\mathrm{W\,m^{-2}}$}			&	\footnotesize{$\mathrm{W\,m^{-1}\,K^{-1}}$}	&	\footnotesize{m}				&	\footnotesize{$\mathrm{W\,m^{-1}\,K^{-1}}$}	&	\footnotesize{K}			&	\footnotesize{K}				&					&					\\ 
		\midrule
		CASE I:				&		$10^4$			&	1					&	$10^{-3}$			&	$2\!\cdot\!10^{-2}$		&	70				&	70				&	0.34				&	0.72				\\
		CASE II:				&		$10^3$			&	$10^{-2}$				&	$10^{-4}$			&	$0.2$				&	$500$	&	3				&	0.17				&	0.14				\\
		MAX($\hat{F}/F_\text{R}$)	&		$520$	&	8					&	$1.1  \!\cdot\! 10^{-3}$	&	$10^{-3}$				&	$1500$	&	$250$	&	2.7				&	1.2				\\
		MIN($\hat{F}/F_\text{R}$)	&		$4  \!\cdot\! 10^4$	&	$10^{-3}$				&	$1.1  \!\cdot\! 10^{-3}$	&	$2  \!\cdot\! 10^{-2}$		&	10				&	$1500$	&	$2.7  \!\cdot\! 10^{-5}$	&	0.28				\\
		MAX($\hat{p}/p_\text{R}$)	&		$6900$	&	8					&	$1.1  \!\cdot\! 10^{-3}$	&	$10^{-3}$				&	$1500$	&	1				&	1.3				&	2.4				\\
		MIN($\hat{p}/p_\text{R}$)	&		10				&	$10^{-3}$				&	$8.7  \!\cdot\! 10^{-6}$	&	$2  \!\cdot\! 10^{-2}$		&	10				&	$1500$	&	0.22				&	$2  \!\cdot\! 10^{-3}$	\\
		 \bottomrule
	\end{tabular}
\end{table*}


As for high $I$ the temperature-dependence of $k_\gas$ and $\eta_\text{dyn}$ can be important, the mean temperature \eqref{eq:A0_co} can be iteratively calculated if $k_\gas=k_\gas(T)$.
Then, the force in the \textit{co} limit $F^\co=F^\co(k(\overline{T}),k_\gas(\overline{T}),\eta_\text{dyn}(\overline{T}))$ with $\overline{T}=A_0^\co$ can be obtained.
In the \textit{fm} regime, the mean temperature \eqref{eq:A0_fm} determines the heat conductivity $k=k(\overline{T}=A_0^\fm)$, and therefore the force.

\section{CONCLUSION}
In the model introduced in Paper 1 (\citep{paper1}) as well as here we incorporate possible temperature differences between the illuminated object and the surrounding gas.
This also includes the case of higher radiative fluxes $I$.
The solutions for the free molecule regime (Paper 1) and the slip flow regime (\eqref{eq:Fphot_low_Knudsen_all}) can be calculated using the given formulae.
The usage of the interpolation between the \textit{fm} and \textit{co} regimes is more complicated. The basic approximation in this paper follows (for simplicity, omitting $h$)
\begin{align*}
   F_\text{phot} &= \frac{2 \, \hat{F}}{\frac{p}{\hat{p}}+\frac{\hat{p}}{p}} \label{eq:Fphot_Transition_RDY} \\
   \hat{F} &= \frac{\pi}{2}\,\sqrt{\frac{\pi}{3}}\,\frac{v_\text{th}\,\eta_\text{dyn}}{\sqrt{A_0^\co\sqrt{T_\infty\overline{T_\gas^\+}}}}\,\sqrt{\tau^\text{co}\,\tau^\text{fm}}\,r_0 \, J_1\, I  \\
   \hat{p} &= \sqrt{\frac{\tau^\text{co}}{\tau^\text{fm}}}\,\frac12\sqrt{3\pi}\,\frac{v_\text{th}\,\eta_\text{dyn}}{r_0} 
\end{align*}
with the mean thermal speed of the gas
\begin{equation*}
   v_\text{th} = \sqrt{\frac{8p}{\pi\rho}}
\end{equation*}
and the mean temperatures
\begin{align*}
	 A_0^\co &= \frac{I\,J_0 + \frac{k_\gas}{r_0}\,T_\infty + \sigma_\text{SB}\varepsilon\left(3\,T_\bb^4+T_\rad^4\right)}{\frac{k_\gas}{r_0} + 4\sigma_\text{SB}\varepsilon\,T_\bb^3}  \\
	 \overline{T_\gas^\+} &= T_\infty+\alpha\left(\frac{I\,J_0 +  \sigma_\text{SB}\varepsilon\,\left(3\,T_\bb^4+T_\rad^4\right)}{4\sigma_\text{SB}\varepsilon\,T_\bb^3}-T_\infty\right) 
\end{align*}
and the scaling factors
\begin{align*}
	 \tau^\text{fm} &= \frac{\sqrt{A_0^\co}}{\sqrt[4]{T_\infty\overline{T_\gas^\+}}} \, \frac{\alpha\,\alpha_\text{m}}{2} \, \frac{1}{\frac{k}{r_0}+ 4\sigma_\text{SB}\varepsilon\,T_\bb^3 }\\
	 \tau^\text{co} &= \frac{\sqrt[4]{T_\infty\overline{T_\gas^\+}}}{\sqrt{A_0^\co}}\,\kappa_\text{s} \, \frac{1}{\frac{k}{r_0}+2\frac{k_\gas}{r_0}+ 4\sigma_\text{SB}\varepsilon\,T_\bb^3 } \; .
\end{align*}
The importance of this model considering strong temperature deviations and high intensities for longitudinal photophoresis becomes apparent when calculating drift motion of dust particles in a (pre-)transitional protoplanetary disk, where the mean free path of the gas is often in the same order as the particles diameters. Especially near the central star the temperatures of the illuminated particles can get significantly higher than the temperature of the surrounding gas. Since photophoresis can dominate the force balance for small particles, the accuracy of the approximation used is highly important and therefore the model given in this paper has to be favored.
Also, particles illuminated with lasers \citep{Daun2008c, Loesche2014} can lead to rather extreme conditions, previously not supported by approximations for the \textit{fm} and transition regimes.


\section{ACKNOWLEDGMENTS}
C.L. was funded by DFG 1385. T.H. was funded by the DFG under the grant number WU321/12-1.

\appendix
\renewcommand{\theequation}{A.\arabic{equation}}
\renewcommand{\thefigure}{A.\arabic{figure}}
\renewcommand{\thesection}{\Alph{section}}
\renewcommand{\thesubsection}{\thesection.\arabic{subsection}}


\section{SUPPLEMENTARIES}\label{sec:supp}
\subsection{One orthogonality relation for associated Legendre polynomials}
\begin{align}
	\int\limits_{-1}^{-1} P_\nu^\mu(x)\,P_\psi^\mu(x)\dd x &= \delta_{\nu\psi}\frac{2}{2 \nu+1}\frac{(\nu+\mu)!}{(\nu-\mu)!} \label{eq:orthogonality}
\end{align}

\subsection{Average}
The mean temperature of the scattered gas $T_\gas^\+$ (\textit{fm}, see Paper 1) is (with $\alpha$ denoting the thermal accommodation coefficient)
\begin{equation}
\overline{T_\gas^\+} = T_\infty+\alpha\left(\overline{T}-T_\infty\right) \; . \label{eq:mean_T_plus}
\end{equation}

\subsection{Transport numbers}
\begin{align}
	\Pe &= \Rey\,\Pra = \frac{\rho\,c_p\,u\,l}{k}   \label{eq:Pe} \\
	\Rey &= \frac{\rho\,u\,l}{\eta_\text{dyn}}   \label{eq:Rey} \\
	\Pra &= \frac{c_p}{k}\eta_\text{dyn}   \label{eq:Pra}
\end{align}

\begin{figure}[!ht]
	\centering
	\includegraphics[width=\columnwidth]{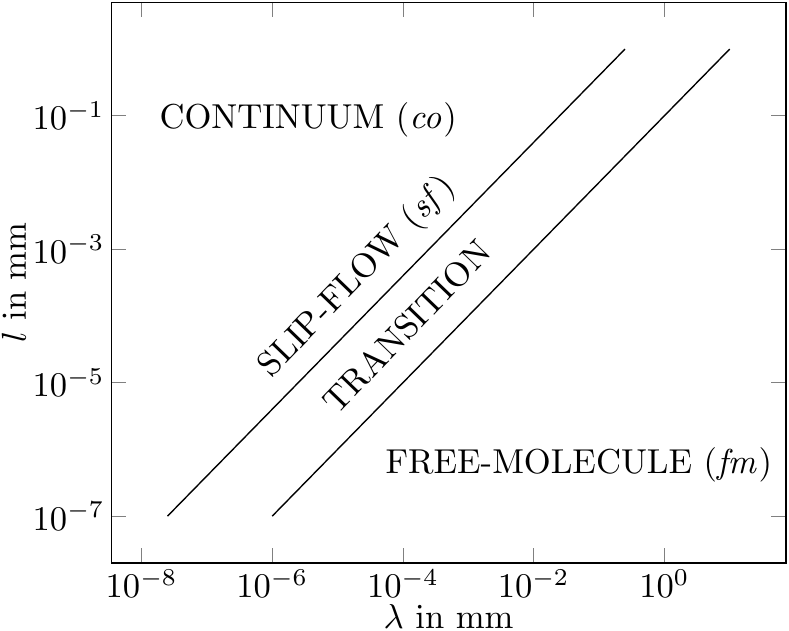}
	\caption{\label{fig:regimes}Knudsen regimes. The Knudsen number is defined as $\Kn = \lambda \, l^{-1}$.}
\end{figure}

\subsection{Force}
The force exerted onto the suspended particle is given by
\begin{subequations}
\begin{align}
   \mathbf{F} &= \int\limits_{\partial V} \underline{\boldsymbol{\Pi}}\cdot \dd\mathbf{A}
   \label{eq:force_Integral} \\
   \underline{\boldsymbol{\Pi}} &= -\rho\,\mathbf{v}\otimes\mathbf{v}+\underline{\boldsymbol{\sigma}} \; . \label{eq:pi}
\end{align}
\end{subequations}
Here, due to the symmetry of the problem, only $F_z$ is not zero.
As it is $\mathrm{d}\mathbf{A}\equiv\mathbf{n}\,\mathrm{d}A$ where $\mathbf{n}=\mathbf{e}_r$ is the normal vector, the product $\underline{\boldsymbol{\Pi}}\cdot\mathbf{n}$ has to be determined
\begin{subequations}
\begin{align}
\underline{\boldsymbol{\Pi}}\cdot\mathbf{n} &= \underline{\boldsymbol{\Pi}}\cdot\mathbf{e}_r \\
&=\left(
\begin{array}{ccc}
\Pi_{rr} & \Pi_{r\zeta} & \Pi_{r\xi} \\
\Pi_{\zeta r} & \Pi_{\zeta\zeta} & \Pi_{\zeta\xi} \\
\Pi_{\xi r} & \Pi_{\xi\zeta} & \Pi_{\xi\xi} \\
\end{array}
\right) \cdot \spaltenvektor{c}{ 1 \\ 0 \\ 0 } \\
&= \spaltenvektor{c}{ \Pi_{rr} \\ \Pi_{\zeta r} \\ \Pi_{\xi r} } = \Pi_{rr}\mathbf{e}_r +  \Pi_{\zeta r}\mathbf{e}_\zeta +  \Pi_{\xi r}\mathbf{e}_\xi \; . \label{eq:Pin}
\end{align}
\label{eq:pi-product}
\end{subequations}
As $\underline{\boldsymbol{\Pi}}$ is given by Eqs. \ref{eq:pi} and \ref{eq:sigma}, the respective parts in spherical coordinates, and with incompressibility are
\begin{subequations}
	\begin{align}
	R_{rr} &= 2\eta_\text{dyn}\frac{\partial v_r}{\partial r} \label{eq:Rrr} \\
	R_{\zeta r} &= \eta_\text{dyn}\left(\frac1{r}\frac{\partial v_r}{\partial\zeta}+r\frac{\partial}{\partial r}\left(\frac{v_\zeta}{r}\right)\right) \label{eq:Rzetar2} \\
	\Pi_{rr} &\stackrel{\eqref{eq:Rrr}}{=} -\rho\,v_r^2-p+2\eta_\text{dyn}\partial_r v_r \\
	\Pi_{\zeta r} &\stackrel{\eqref{eq:Rzetar}}{=} -\rho\,v_\zeta\,v_r+\eta_\text{dyn}\left(\frac1{r}\partial_\zeta v_r + r\,\partial r\left(\frac{v_\zeta}{r}\right)\right) \\
	\Pi_{\xi r} &= 0 \; .
	\end{align}
	\label{eq:R_zwei}
\end{subequations}
$\Pi_{\xi r} = 0$ as $v_\xi=0$, and $\mathbf{v}$ is independent of $\xi$.
The $z$-component of the product in \eqref{eq:pi-product} is
\begin{equation}
   \left(\underline{\boldsymbol{\Pi}}\cdot\mathbf{n}\right)_z = \Pi_{rr}\cos\zeta - \Pi_{\zeta r}\sin\zeta \; ,
\end{equation}
and therefore the $z$-component of the force (\eqref{eq:force_Integral}) reads
\begin{align}
   F_z &= 2\pi\,r_0^2\int\limits_{0}^{\pi}\mathrm{d}\zeta\,\sin\zeta\left( \Pi_{rr}\cos\zeta - \Pi_{\zeta r}\sin\zeta \right) \; . \label{eq:Fz2}
\end{align}
\begin{table}[!h]
	\caption{\label{tab:notation}Notation.}
\end{table}\vspace*{-5mm}
\begin{supertabular}{l p{0.8\columnwidth}}
\toprule	
\text{variable} & meaning 	\\
\midrule		
$\mathbf{r} = (r,\zeta,\xi)$ & spherical coordinates (\fig{fig:sphere_with_Trot}) \\
$r_0$				& radius of spherical particle suspended in gas \\
$\mathbf{n},\, \mathbf{t}$		& normal and tangent vectors of a surface \\
$\partial V$ & border of the volume V, i.e. $r=r_0$ for the sphere \\
$\mathbf{v}$ & gas mass velocity in $\mathrm{m\,s^{-1}}$ \\
$v_\text{th}$ & mean thermal gas speed \\
$\mathbf{u}$ & velocity of the suspended particle, relative to the gas \\
$T(r,\zeta,\xi)$		& particle temperature in $\mathrm{K}$	\\
$\overline{T}, \widetilde{T}$ & mean particle surface temperatures in $\mathrm{K}$ (Eqs. \ref{eq:mean_temperature} and \ref{eq:mean_temperature4}) \\
$T_\gas$ & gas temperature \\
$T_\infty$ & gas temperature far away from the particle \\
$T_\gas^{\oplus/\ominus}$	& gas temperature for velocity half-spaces $\mathbf{n}\cdot\mathbf{v}>0$ and $\mathbf{n}\cdot\mathbf{v}<0$ (\fig{fig:sphere_with_Trot}), used in the \textit{fm} regime (see Paper 1), here we write $T_\gas^{\ominus}=T_\infty$	\\
$\overline{\left.T_\gas\right|_{\partial V}}$ & mean temperature of the gas layer around the particle \\
$T_\rad$	& temperature of external radiation field	\\
$T_\text{bb}$ & black-body temperature (\eqref{eq:blackBodyTemp}) \\
$R_\gas$ & universal gas constant in $\mathrm{J\,mol^{-1}\,K^{-1}}$ \\
$M$ & molar gas mass in $\mathrm{kg\,mol^{-1}}$ \\
$p$				& gas pressure in $\mathrm{Pa}$ \\
$\hat{p}$ & gas pressure where $\mathbf{F}_\text{phot}$ maximizes (\eqref{eq:p_max})\\
$p^*$ & characteristic gas pressure (\eqref{eq:char_p})\\
$\rho$			& gas mass density in $\mathrm{kg\,m^{-3}}$ \\
$\underline{\boldsymbol{\sigma}},\,\underline{\boldsymbol{R}}$ & stress and friction tensor (\eqref{eq:sigma}) \\
$\psi$ & stream function \\
$E^2$ & stream function operator (\eqref{eq:stream_function_operator})\\
$P_\nu^\mu$ & associated Legendre polynomial \\
$\mathbf{F}_\text{phot}$	& photophoretic force \\
$\hat{F}$	& maximum photophoretic force at a pressure $\hat{p}$ (\eqref{eq:F_max}) \\
$\delta$ & stretch factor in \eqref{eq:Fphot_Transition} \\
$\tau^{\fm},\, \tau^{\co}$ & dimensionless scaling coefficients for $\hat{F}$ and $\hat{p}$ (\eqref{eq:taus}) \\
$\Xi$ & scaling constant for $\hat{F}$ and $\hat{p}$ in $\mathrm{Pa\,m\,K^{-1}}$ (\eqref{eq:Xi}) \\
$\varphi_\text{rad},\, \vartheta_\text{rad}$ & dimensionless solution numbers (\eqref{eq:ht_var})\\
$\alpha,\, \alpha_\text{m}$ & thermal and momentum accommodation coefficient (dimensionless) \\
$\kappa_\text{t}$ & temperature jump coefficient (dimensionless), related to $\alpha$ (\eqref{eq:temperature_jump_coeff}) \\
$\kappa_\text{m}$ & gas-kinetic frictional slip (or momentum exchange) coefficient (dimensionless), related to $\alpha_\text{m}$ \\
$\kappa_\text{h}$ & thermal stress slip coefficient (dimensionless) \\
$\kappa_\text{s}$ & thermal creep (or thermal slip) coefficient (dimensionless), related to $\alpha_\text{m}$ (\eqref{eq:thermal_slip_coeff}) \\
$J_\nu$ & asymmetry factor (dimensionless, \eqref{eq:asymmetry_factor_all}) \\
k				& thermal conductivity of suspended particle in $\mathrm{W\,m^{-1}\,K^{-1}}$	\\
$k_\gas$			& thermal conductivity of the gas	\\
$\eta_\text{kin},\, \eta_\text{dyn}$ & kinematic and dynamic viscosity, $\eta_\text{kin} = \eta_\text{dyn}/\rho$, $\eta_\text{dyn}$ in $\mathrm{Pa\,s}$ \\
$\Pe$			& P{\'e}clet number (\eqref{eq:Pe}) \\
$\Rey$ 			& Reynolds number (\eqref{eq:Rey}) \\
$\Pra$ 			& Prandtl number   (\eqref{eq:Pra}) \\
$h$		& heat transfer coefficient (\eqref{eq:h_fm}) in $\mathrm{W\,m^{-2}\,K^{-1}}$ \\
$I$				& effective intensity $I=\varepsilon\, I_0$ in $\mathrm{W\,m^{-1}}$ \\
$\varepsilon$		& (mean) emissivity \\
$\sigma_\text{SB}$ & Stefan-Boltzmann constant in $\mathrm{W\,m^{-2}\,K^{-4}}$ \\
$\lambda$ & mean free path of the gas in $\mathrm{m}$ \\
$\Kn$ & Knudsen number (dimensionless, \eqref{eq:Knudsenzahl}) \\
$q$				& normalized source function (\eqref{eq:heatEQ}) in $\mathrm{m^{-1}}$ \\
$A_\nu,\,B_\nu,\,C_\nu,\,q_\nu$ & expansion coefficients ($\nu\ge 0$) \\		
\bottomrule
\end{supertabular}

\FloatBarrier
\section*{References}






\bibliographystyle{model5-names}\biboptions{authoryear}


\bibliography{references}

\end{document}